\documentclass[sigconf, nonacm]{acmart}

\usepackage{balance} % recommended to use by VLDB formating guidelines
\usepackage{array} %This package is used for fixing the issue with table. The regular package does not allow for center alignment and sizing of table appropriately. 
\usepackage{graphicx} %for graphics
\usepackage[labelfont=bf,textfont=bf,font=small]{subcaption} %%for sub captions - usually helpful in listings or subfigs
\usepackage[labelfont=bf,textfont=bf,font=small]{caption} % for captions. 
\usepackage{pifont} %\ding is part of this package. for markings
\usepackage{multirow}
\usepackage[ruled,vlined]{algorithm2e}
\usepackage{listings}
\usepackage{comment}
\usepackage{pifont}% http://ctan.org/pkg/pifont
\usepackage[normalem]{ulem} %only needed for drafting
\usepackage{enumitem}
\usepackage[a-1b]{pdfx}

\DeclareCaptionFormat{mylst}{\hrule height 1pt\vspace{2pt}#1#2#3}
\newcommand{\pname}[1]{{{{EMOGI}}}{#1}}

\newcommand\vldbdoi{10.14778/3425879.3425883}
\newcommand\vldbpages{114-127}
\newcommand\vldbvolume{14}
\newcommand\vldbissue{2}
\newcommand\vldbyear{2021}
\newcommand\vldbauthors{\authors}
\newcommand\vldbtitle{\shorttitle} 
\newcommand\vldbavailabilityurl{https://github.com/illinois-impact/EMOGI}

\begin{document}
\title{EMOGI: Efficient Memory-access for Out-of-memory Graph-traversal in GPUs}
\pagestyle{empty}
%%
%% The "author" command and its associated commands are used to define the authors and their affiliations.
\author{Seung Won Min}
\affiliation{%
  \institution{University of Illinois at Urbana-Champaign}
  \city{Urbana}
  \state{IL}
  \country{USA}
%  \institution{University of Illinois at Urbana-Champaign \& \\IBM-Illinois C3SR Center}
}
\email{min16@illinois.edu}

\author{Vikram Sharma Mailthody}
\affiliation{%
  \institution{University of Illinois at Urbana-Champaign}
  \city{Urbana}
  \state{IL}
  \country{USA}
%  \institution{University of Illinois at Urbana-Champaign \& \\IBM-Illinois C3SR Center}
}
\email{vsm2@illinois.edu}

\author{Zaid Qureshi}
\affiliation{%
  \institution{University of Illinois at Urbana-Champaign}
  \city{Urbana}
  \state{IL}
  \country{USA}
%  \institution{University of Illinois at Urbana-Champaign \& \\IBM-Illinois C3SR Center}
}
\email{zaidq2@illinois.edu}

\author{Jinjun Xiong}
\affiliation{%
  \institution{IBM T.J. Watson Research Center}
  \city{Yorktown Heights}
  \state{NY}
  \country{USA}
%  \institution{IBM T.J. Watson Research Center \& \\IBM-Illinois C3SR Center}
}
\email{jinjun@us.ibm.com}

\author{Eiman Ebrahimi}
\affiliation{%
  \institution{NVIDIA}
  \city{Austin}
  \state{TX}
  \country{USA}
}
\email{eebrahimi@nvidia.com}

\author{Wen-mei Hwu}
\affiliation{%
  \institution{University of Illinois at Urbana-Champaign}
  \city{Urbana}
  \state{IL}
  \country{USA}
%  \institution{University of Illinois at Urbana-Champaign \& \\IBM-Illinois C3SR Center}
}
\email{w-hwu@illinois.edu}

\begin{abstract}

{Modern analytics and recommendation systems are increasingly based on graph data that capture the relations between entities being analyzed.} Practical graphs come in huge sizes, offer massive parallelism, and are stored in sparse-matrix formats such as compressed sparse row (CSR). 
To exploit the massive parallelism, developers are increasingly interested in using GPUs for graph %analytics and 
traversal. However, due to their sizes, graphs often do not fit into the GPU memory.
Prior works have either used input data pre-processing/partitioning or unified virtual memory (UVM) to 
%automatically 
migrate chunks of data from the host memory to the GPU memory.
However, the large, multi-dimensional, and sparse nature of graph data presents a major challenge to these schemes and 
%result in low data spatial locality, 
results in significant amplification of data movement and reduced effective data throughput.
In this work, we propose \pname{}, an alternative approach to traverse graphs that do not fit in GPU memory using direct cache-line-sized access to data stored in host memory. 
%

%EIMAN-5-31-20: making some changes to this opening sentence, but the previous sentence is commented below it
This paper addresses the open question of whether a sufficiently large number of overlapping cache-line-sized accesses can be sustained to 1) tolerate the long latency to host memory, 2) fully utilize the available bandwidth, and 3) achieve favorable execution performance. 
%
%EIMAN-5-31-20: I think you could be fine without including the following sentence in abstract. Right now abstract feels too long for my liking.
We analyze the data access patterns of several graph traversal applications in GPU over PCIe using an FPGA to understand the cause of poor external bandwidth utilization. 
By carefully coalescing and aligning external memory requests, we show that we can minimize the number of PCIe transactions and nearly fully utilize the PCIe bandwidth %even 
with direct cache-line accesses to the host memory.
%
%EIMAN-5-31-20: you can probably safely lose the following statement from abstract also.
%EIMAN-5-31-20: ``direct host memory access'' in sentence below?
%Even though \pname{}'s direct access removes some of the opportunity of exploiting temporal locality, we find the performance %improvements obtained by direct access 
%is higher than 
%the overheads introduced by 
%pre-processing or runtime data management.  
%
%EIMAN-5-31-20: if you're only stating one number in abstract, stating an average number is better than stating an "up to number" imo. stating an 'up to' number doesn't give a good feel for the generality of your approach which is what the reader probably cares most about here.
%As a result,
\pname{} achieves 2.60$\times$ speedup on average compared to the optimized UVM implementations in various graph traversal applications.
%As a result, \pname{} achieves up to 6.34$\times$ speedup compared to the baseline UVM implementations of %various graph traversal applications such as
%breadth-first search, single-source shortest paths, and connected components when operating on a variety of large graphs.
%
%EIMAN-5-31-20: re-wrote the sentence below and left old one there commented
We also show that \pname{} scales better than a UVM-based solution when the system uses higher bandwidth interconnects such as PCIe 4.0.

\end{abstract}

\maketitle

\begingroup\small\noindent\raggedright\textbf{PVLDB Reference Format:}\\
\vldbauthors. \vldbtitle. PVLDB, \vldbvolume(\vldbissue): \vldbpages, \vldbyear.\\
\href{https://doi.org/\vldbdoi}{doi:\vldbdoi}
\endgroup
\begingroup
\renewcommand\thefootnote{}\footnote{\noindent
This work is licensed under the Creative Commons BY-NC-ND 4.0 International License. Visit \url{https://creativecommons.org/licenses/by-nc-nd/4.0/} to view a copy of this license. For any use beyond those covered by this license, obtain permission by emailing \href{mailto:info@vldb.org}{info@vldb.org}. Copyright is held by the owner/author(s). Publication rights licensed to the VLDB Endowment. \\
\raggedright Proceedings of the VLDB Endowment, Vol. \vldbvolume, No. \vldbissue\ %
ISSN 2150-8097. \\
\href{https://doi.org/\vldbdoi}{doi:\vldbdoi} \\
}\addtocounter{footnote}{-1}\endgroup

\ifdefempty{\vldbavailabilityurl}{}{
\vspace{.3cm}
\begingroup\small\noindent\raggedright\textbf{PVLDB Availability Tag:}\\
The source code of this research paper has been made publicly available at \url{\vldbavailabilityurl}.
\endgroup
}

\section{Introduction}
Graph workloads are becoming increasingly widespread and common in various applications such as social network analysis, recommendation systems, financial modeling, bio-medical applications, graph database systems, web data, geographical maps, and many more~\cite{scalegraphissue, suitesparse,ubicrawler,networkrepo, Friendster, GAP, BoVWFI,BRSLLP,BCSU3}. 
Graphs used in these applications often come in huge sizes. 
A recent survey conducted by the University of Waterloo~\cite{scalegraphissue} finds that many organizations use graphs that consist of billions of edges and consume hundreds of gigabytes of storage. 

The main challenge that graph application developers currently face is performing graph traversal computations on large graphs~\cite{scalegraphissue}. 
Because of the massive parallelism present in the graph traversal computation, GPUs are increasingly used to perform graph analytics. 
However, the ability to process large graphs in GPUs is currently hampered by their limited memory capacity. 
Thus in this work, we primarily focus on developing an efficient graph traversal system using  GPUs that accesses large graph data from host memory. 

For efficient storage and access, graphs are stored in a compressed sparse row (CSR) data format as it has low memory overhead. In CSR format, a graph is stored as the combination of a vertex list and an edge list.
Even with CSR data format, large graph datasets cannot fit in today's GPU memory. 
Thus, most prior works store these large graphs in host memory and have GPUs access them through the unified virtual memory (UVM) mechanism~\cite{uvm, Gera20, etc, batchaware, graphchallenge18, pageplacement, mosaic, gpuswap}.
UVM brings both CPU memory and GPU memory into a single shared address space.
UVM allows GPUs to simply access the data in the unified virtual address space and it transparently migrates required pages between host memory and GPU memory using a paging mechanism. 

However, several prior work~\cite{Gera20, etc, batchaware, graphchallenge18, pageplacement, mosaic, gpuswap} have reported that the performance of graph traversal using UVM is not competitive.
This is because memory accesses that go to the edge list during graph traversal are irregular in nature. 
Furthermore, based on our analysis of 1122 graphs that have at least 1M vertices and edges from LAW~\cite{ubicrawler}, SuiteSparse Matrix Collection~\cite{suitesparse}, and  Network Repository~\cite{networkrepo}, we find the average degree per vertex is about 71.
This implies that when those graphs are represented in a compressed adjacency list format such as CSR, each vertex's neighbor edge list is about 71 elements long on average.
Thus transferring an entire 4KB page, as in the case of UVM, can cause memory thrashing and unnecessary I/O read amplification. 

As a result, prior works have proposed pre-processing of input graphs by partitioning and loading those edges that are needed during the computation~\cite{gunrock,graphreduce,graphie,Sabet20} or proposing UVM specific hardware or software changes such as locality enhancing graph reordering~\cite{Gera20}, GPU memory throttling~\cite{etc,batchaware}, overlapping compute and I/O~\cite{chai}, or even proposing new prefetching policies in hardware that can increase data locality in GPU memory~\cite{pageplacement, mosaic, gpuswap}.

In this work, we take a step back and revisit the existing hardware memory management mechanism for when data does not fit in GPU memory.
Specifically, we focus on zero-copy memory access which allows GPUs to directly access the host memory in cache-line granularity.
With zero-copy, no complicated data migration is needed and GPUs can fetch data as small as 32-byte from the host memory.
Even with such advantages, unfortunately, zero-copy is known to have underwhelming performance due to the low external interconnect bandwidth~\cite{gangulyadaptive}.
Interestingly, however, we do not find any systematic analysis showing the exact limiting factor of the zero-copy performance or leading to any effort to improve it.

Instead of making a premature conclusion, we build a system with a custom-designed FPGA-based PCIe traffic monitor and explore any opportunity to optimize zero-copy performance.
We use the system to address the question of whether a sufficiently large number of overlapping cache-line-sized accesses can be sustained to 1) tolerate the long latency to host memory, 2) fully utilize the available bandwidth, and 3) achieve favorable execution performance for graph traversal applications. 
To this end, the key goal of our work is to avoid performing any pre-processing or data manipulation on the input graph and allowing GPU threads to directly perform cache-line-sized accesses to data stored in host memory during graph traversals.

By using a toy example, we show that by naively enabling zero-copy, the system
cannot saturate the PCIe 3.0 x16 bandwidth (see $\S$~\ref{sec:zerocopy.opt}). 
To address this, we propose two key software optimizations needed to best exploit PCIe bandwidth for the zero-copy access. 
First, we propose the merged memory access optimization that optimizes for generating maximum-sized PCIe request to zero-copy memory (see $\S$~\ref{sec:zerocopy.opt}). 
Second, we propose forcing memory access alignment by shifting all warps to 128-byte boundaries when there is misalignment.
This is because the memory access merge optimization does not guarantee memory request alignment. Such misalignment can result in performance degradation.
While these optimizations sacrifice some parallelism and incur additional control divergence during kernel execution, their benefit in terms of improved bandwidth utilization far outweighs the cost.
We then apply these two optimizations to popular graph traversal applications
%\footnote{The source codes are available at \href{https://github.com/illinois-impact/EMOGI}{\textit{\color{cyan}{https://github.com/illinois-impact/EMOGI}}}.}
including breadth-first search (BFS), single-source shortest path (SSSP), connected components (CC), and PageRank (PR) to enable efficient traversal on large graphs. 

Using real-world and synthetic large graphs (see Table~\ref{tab:dataset}), we show that \pname{} can achieve 2.93$\times$ speedup on average compared to the optimized UVM implementations of BFS, SSSP, CC, and PR benchmarks across a variety of graphs. 
We also evaluate \pname{} on the latest generation of the NVIDIA Ampere A100 GPU with PCIe 4.0 and show that \pname{} still remains performant and scales better than the UVM solution when using higher-bandwidth interconnect. 
\pname{} achieves speedups of up to 4.73$\times$ over current state-of-art GPU solutions~\cite{Gera20, Sabet20} for large out-of-memory graph traversals.
In addition, \pname{} does not require pre-processing or runtime page migration engine. 

To the best of our knowledge, \pname{} is the first work to systematically characterize GPU PCIe access patterns to optimize zero-copy access and to provide in-depth profiling results of varying PCIe access behaviors for a wide range of graph traversal applications.
Overall, our main contributions can be summarized as follows: 
\begin{enumerate}
    \item We propose \pname{}, a novel zero-copy based system for very large graph traversal on GPUs.
    \item We propose two zero-copy optimizations, memory access merge and memory access alignment, that can be applied to graph traversal kernel code to maximize PCIe bandwidth.
    \item We show \pname{} performance scales linearly with CPU-GPU interconnect bandwidth improvement by evaluating PCIe 3.0 and PCIe 4.0 interconnects.
\end{enumerate}

The rest of the paper is organized as follows:
we provide a brief primer on GPU based graph traversal and the challenges in executing graph traversals using UVM in $\S$~\ref{sec:background}.
We then discuss how to enable zero-copy memory with GPUs and discuss the reasons for its poor performance in a naive but common kernel code pattern in $\S$~\ref{sec:zerocopy}.
Using the gained insights, we then apply zero-copy optimizations to graph traversal algorithms in $\S$~\ref{sec:implementation}.
We discuss \pname{}'s performance improvement for various graph traversal algorithms on several large graphs in $\S$~\ref{sec:eval}. 
$\S$~\ref{sec:relatedwork} discusses how \pname{} differs from prior work and we conclude in $\S$~\ref{sec:conclusion}.
\section{Background}
\label{sec:background}
In this section, we first provide a brief primer on GPU based graph traversal.
Then we will describe techniques used to traverse graphs that cannot fit into the GPU memory.

\subsection{Parallelizing Graph Traversal on GPUs}
\label{subsec:par-graphtraversal}

%Graph traversals can be largely divided into a vertex-centric~\cite{largegraphcuda,cusha,thinkvertexthinkgraph} method and an edge-centric~\cite{xstream} method.
%
%The vertex-centric method can be further divided into a scatter-based method and a gather-based method.
%
%In this paper, we mainly focus on the vertex-centric + scatter method due to its simplicity.
%

The exact workflow of the graph traversal depends on the type of the application and the optimization level, but a general flow can be described with  Algorithm~\ref{algo:traversal}.
First, before the traversal begins, initial active vertices need to be set.
In case of BFS, only a single vertex needs to be set as active, which is basically a source vertex.
Once all the initial active vertices are set, the graph traversal can begin.
Graph traversal is composed of multiple iterations of sub-traversals.
In each sub-traversal, all immediately neighboring vertices of the currently active vertices are exhaustively traversed.
The condition to set the next active vertices depends on the type of application as well.
In case of BFS, any neighboring vertices which are not visited ever before are marked to be the next active vertices.
The traversal ends once there are no more active vertices left in the graph.

\begin{algorithm}[b]
\SetAlgoLined
set\_initial\_active\_vertex()\\
\While{there exist active vertices in G}{
%\While{there exist active node v$_1$ in G}{
\For{\textbf{\textup{all}} vertices v$_1$ in Graph G}{
\If{v$_1$ is active}{
set \textit{v}$_1$ as inactive\\
\For{\textbf{\textup{all}} neighbors v$_2$ of v$_1$}{
application\_dependant\_workload()\\
\If{application\_dependant\_condition()}{
set \textit{v}$_2$ as active
}
}
}
}
}
 \caption{High-level Graph Traversal Flow}
 \label{algo:traversal}
\end{algorithm}

The main benefit of the GPU implementation of the graph traversal comes from the massive number of vertices~\cite{BoVWFI,BRSLLP,Friendster}.
With a help of several atomic instructions, both the inner loop and the outer loop in Algorithm~\ref{algo:traversal} can be fully parallelized with GPU for various kinds of graph traversal applications~\cite{largegraphcuda,hawick10,gascl,maxwarp}.

As an input graph format for the GPU graph traversal, we use compressed sparse row (CSR) format.
CSR is arguably the most popular way to represent a graph because of its low memory overhead~\cite{gunrock,graphreduce,graphie,Sabet20,multigpugraph,medusa}.
Certain graph processing frameworks such as nvGRAPH~\cite{nvGRAPH} support other input formats like coordinate list (COO), but they internally convert the inputs to the CSR format before the actual processing step.
CSR encodes the entire graph with just 2 arrays, as shown in Figure~\ref{fig:csr}. 
The edge list stores each vertex's neighbor list sequentially, such that all the neighbors of vertex 0 are stored first, then the neighbors of vertex 1, and so on. 
The vertex list is indexed by a vertex ID and stores the starting offset of that vertex's neighbor list in the edge list.
The datatypes of both edge and vertex lists can vary depending on the graph size.
For example, using a 4-byte datatype for the edge list can identify at most 4 billion nodes.

\subsection{Out-of-Memory Graph Traversal on GPUs}

\begin{figure}
  \centering
  \includegraphics[width=.8\linewidth]{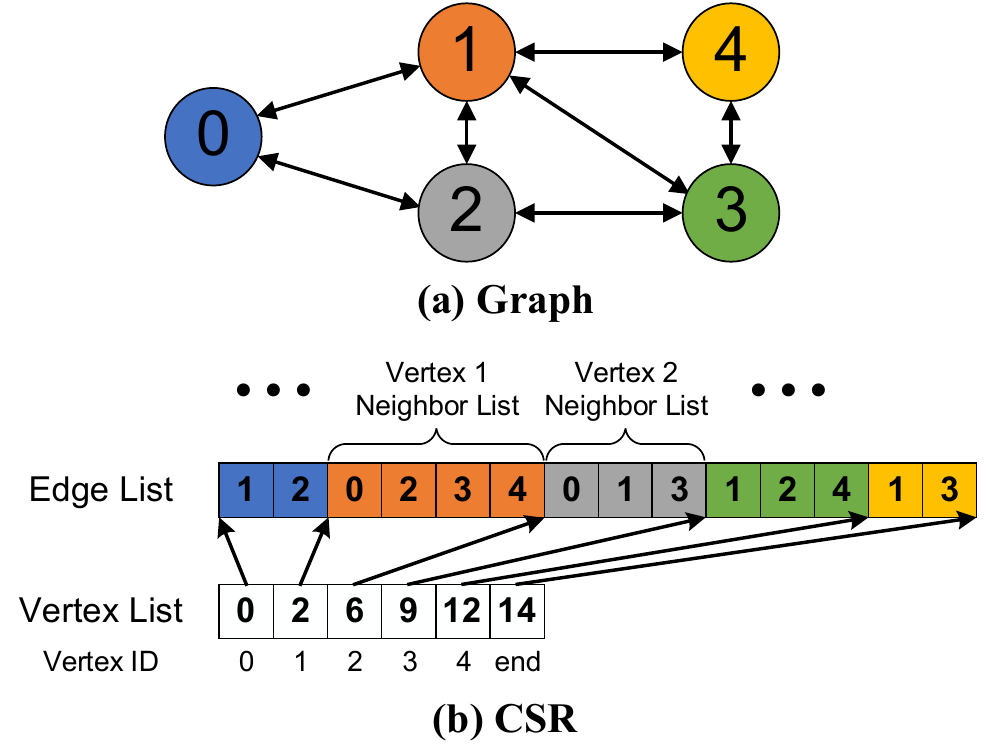}
  \caption{Sample undirected (a) graph and its (b) CSR representation. The edge list contains the neighbor list for each node. The vertex list is indexed by vertex IDs and contains the offsets for the starting position of that vertex's neighbor list in the edge list.}
  \label{fig:csr}
\end{figure}

Graphs, even in the CSR format, 
%are getting 
can be orders of magnitude larger than GPU memory.
The easiest way to enable GPU based graph traversal on such graphs is to use the Unified Virtual Memory (UVM)~\cite{uvm, pascalwhite, voltawhite,Gera20, etc, batchaware, graphchallenge18, graphchallengetc19, pageplacement, mosaic, gpuswap}.
UVM is a unified memory management module that provides a single memory address space accessible by both CPU and GPU through the page faulting mechanism.
UVM reduces the burden on the programmer as they do not have to explicitly manage where the data resides.
UVM transparently allows device memory over-subscription with the use of CPU memory, enabling computation on large data sets that exceed GPU device memory capacity. 
The UVM driver is responsible for on-demand page migration between the CPU and GPU.

The granularity of the data migration may vary depending on the data access pattern, but the minimum granularity is a system page size (4KB).
Once the page is migrated, subsequent accesses to the same page do not need additional data migrations and the accesses can directly go to the GPU memory.
If the memory footprint of the kernel is larger than the GPU memory, some pages need to be evicted from the GPU memory to host other pages during the kernel runtime.
Since the entire management process is single-threaded, the overall performance of the UVM page migration heavily depends on the single-thread performance of the host CPU.

The inefficiency of UVM in graph traversal comes in two ways.
First, for the very large graphs, it is hard to exploit temporal locality as the limited GPU memory capacity will cause frequent page thrashing.
Second, there is a lack of spatial locality between neighbor lists, causing significant I/O read amplification and more frequent page migrations.
For example, in Figure~\ref{fig:csr}, the neighbor lists of the vertex 1 and 3 need to be accessed at the same time we start BFS from the vertex 4.
However, as shown in the CSR representation, the lists are non-contiguous in the edge list.
In a more realistic case with a large graph, these lists can be separated by millions of elements in the edge list.
Therefore, accessing these two lists will likely generate two separate 4KB page migrations.
Assuming that all accesses to the different neighbor lists will generate separate 4KB page migrations, all neighbor lists should have least 512 to 1024 of elements (depends on the datatype size) to make the 4KB data transfer 100\% efficient, which might be quite challenging.
By combining the frequent page migrations caused by the lack of data locality and the high page fault handling overhead of UVM, GPU performance can be severely throttled. 
\section{Zero-Copy}
\label{sec:zerocopy}
%\fixme{Vikram:need to establish in section 2 that cacheline requests is better than page request for graph traversal application. No need to have numbers but a logic explanation is needed. }
%\david{David: Is this a question or a notification} notification :) 
To allow GPU threads access to the external memory in smaller granularity than UVM, GPUs support marking memory address ranges as zero-copy memory~\cite{cudapractice}.
%
%Zero-copy, also often referred to as \emph{direct access}, does not require any page migration or duplication from the external memory to the GPU memory.
Zero-copy, also often referred to as \emph{direct access}, does not require any page migration or duplication between the external and GPU memories.
%
%\fixme{Instead, zero-copy transforms cacheline-sized memory requests from GPU threads to the external interconnect protocol such as PCIe or NVLink.}
%Instead, GPU threads access zero-copy memory as if it was GPU global memory, and the GPU transforms memory requests from the threads to memory requests over the external interconnect protocol such as PCIe or NVLink.
Instead, GPU threads access zero-copy memory as if it was GPU global memory, and the GPU transforms memory requests from the threads to memory requests over an external interconnect like PCIe.
%
%\fixme{The target of the memory requests can be anywhere in the system's memory address space.} %does this convey the same meaning ? 
The target of the memory requests can be anywhere in the system as long as the location can be memory-mapped into the shared bus address.
Common examples include system memory, peer-connected PCIe network interface cards, and peer-connected GPUs.
Due to the high latency of the external interconnects, using zero-copy was thought to have low bandwidth~\cite{gangulyadaptive} and thus used for only accessing small and frequently shared data.
% such as hardware control registers of peer-connected devices. % or software defined locks between CPUs and GPUs.
%\fixme{<-i dont get the point of this sentence here}
%
In this section, we describe how to enable zero-copy and use a peer-connected FPGA to explore any optimization opportunities available for zero-copy in detail.
Based on the analysis, we apply several optimizations and show correctly using zero-copy can nearly saturate the PCIe bandwidth.

\subsection{Enabling Zero-Copy}
\label{sec:zerocopy.enable}

From the system's point of view, zero-copy is enabled as follows:
%
%First, the data which needs to be shared with GPU must be pinned in the host memory.
First, the data to be shared with GPU must be pinned in the host memory.
Pinned memory 
%requires the location of the allocated memory in DRAM so it 
cannot be swapped out to the disk or relocated by the host OS memory manager.
Second, the corresponding bus address (e.g. PCIe) of the pinned data should be mapped into the GPU page table so the GPU can generate a correct external memory request.
Finally, the mapped address should be passed to the user space so the programmer can use pointers in the GPU kernel to access the region.

From CUDA API's point of view, zero-copy can be enabled in three ways. 
First technique uses \texttt{cudaMallocManged()} to allocate UVM space and applies \texttt{cudaMemAdviseSetAccessedBy} flag with \texttt{cudaMemAdvise()}.
The resulting data pointer can be directly used from CUDA kernels to generate zero-copy memory access.
One thing worth noting here is that the \texttt{cudaMemAdviseSetAccessedBy} flag should not be used with other \texttt{cudaMemAdvise()} flags since the other flags override \texttt{cudaMemAdviseSetAccessedBy}.
Second is by using \texttt{cudaMallocHost()}.
This is the simplest method since the memory allocated by \texttt{cudaMallocHost()} can be directly used in the CUDA kernel to do zero-copy access.
The last scheme uses general memory allocators, like \texttt{malloc()}, and \texttt{cudaHostRegister()} and \texttt{cudaGetDevicePointer()} on top of the allocated memory.
In this case, the \texttt{cudaHostRegister()} pins the allocated memory space and \texttt{cudaGetDevicePointer()} returns a CUDA-compatible pointer.
Our experiments showed all three techniques provided the same performance. 
%\fixme{We need a sentence here to tell that all three has very similar or same behaviors when it comes to accessing the memory. I am more inclined to similar over same. We need it address an obvious review question on next para } \david{David: Is this a question or a notification}

\begin{figure}[t]
  \centering
  \includegraphics[width=\linewidth]{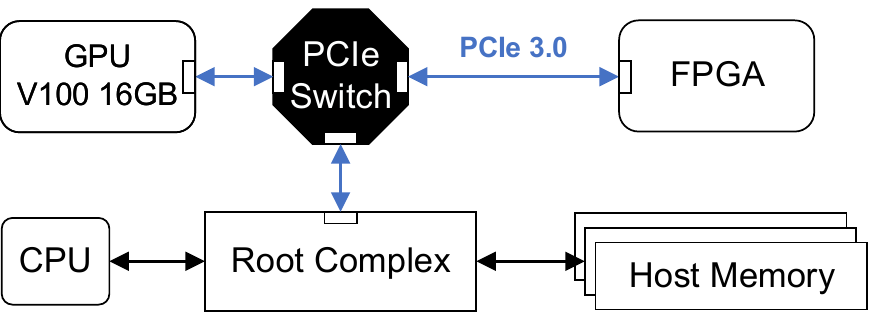}
  \caption{PCIe traffic monitoring environment. The FPGA is used to characterize the zero-copy memory access pattern from GPU.}
  \label{fig:setup}
\end{figure}

\begin{figure*}[t]
  \centering
  \includegraphics[width=\textwidth]{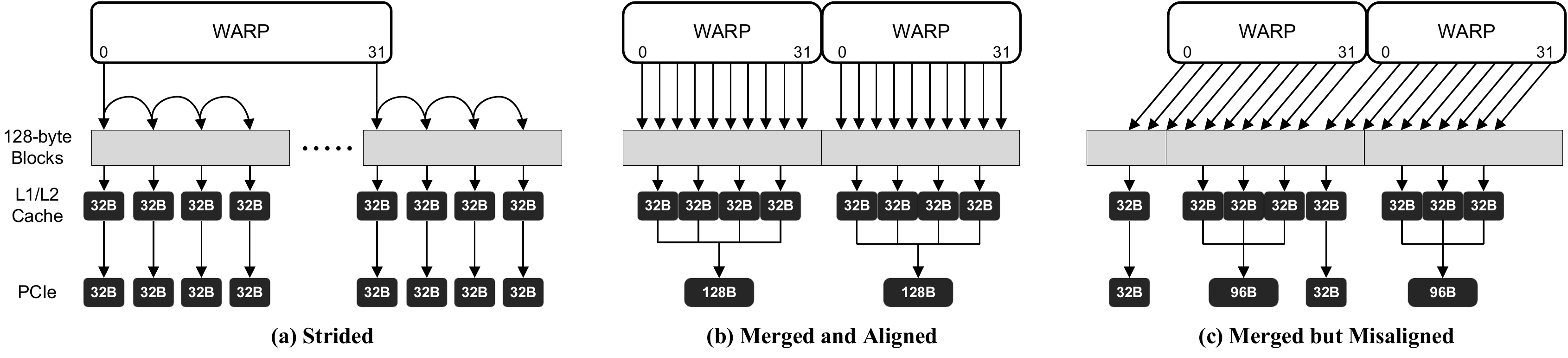}
  \caption{GPU PCIe memory request patterns observed with FPGA. In (a), each thread scans a different 128B block and end up making multiple 32B PCIe memory read requests. In (b), individual 32B memory read requests in a contiguous address space occur at the same time and GPU merges them into a single 128B PCIe memory read request. In (c) each warp is performing a misaligned memory request (off by 32B from 128B boundary) resulting in generating a 32B PCIe and 96B PCIe request. In this figure, we assume each memory access is 4B.}
  %\caption{GPU PCIe memory request patterns observed with FPGA. In (a), memory requests which could have been merged into a larger memory request are sent as multiple separate 32B PCIe memory read requests. In (b), individual 32B memory read requests in a contiguous address space occurs at the same time and GPU merges them into a single 128B PCIe memory read request. In (c) each warp is performing a misaligned memory request (off by 32-byte from 128-byte boundary) resulting in generating a 32-byte PCIe and 96-byte PCIe request. In this figure, we assume each memory access size is 4B.}
  \label{fig:coalescing}
\end{figure*}

\subsection{Zero-Copy Analysis Setup}
\label{sec:eval.setup.fpga}
%\label{sec:eval.setup.fpga}

In order to understand how GPU accesses external zero-copy memory over PCIe, we designed and built the monitoring system shown in Figure~\ref{fig:setup}. 
The FPGA is connected to the GPU using a PCIe switch in peer-to-peer mode. 
Furthermore, the FPGA is programmed to advertise itself as a large memory using the base address register (BAR) region provided by the PCIe specification~\cite{pciespec}.
This advertised FPGA memory region can be mapped to the user space using the \texttt{mmap()} system call.
The returned pointer value from the \texttt{mmap()} call can be used by CPU to access the FPGA as a zero-copy region.
To allow the GPU direct access to the FPGA, we pass the pointer to \texttt{cudaHostRegister()} and \texttt{cudaGetDevicePointer()} CUDA APIs.
The final pointer generated by the two APIs can be passed to the CUDA kernel code and dereferenced by GPU threads, thus allowing zero-copy access 
%as described in $\S$~\ref{sec:zerocopy.enable} 
to the FPGA. 
Using this system, we can now analyze the low-level PCIe traffic of zero-copy memory access by the GPU. 
%\fixme{Now that we understand how to enable Zero-copy memory, we next need to understand how the zero-copy memory access pattern behaves when accessing external memory using PCIe (or NVLink).}%to fix
%
%To identify low-level PCIe traffic characteristics of zero-copy, we use a peer-to-peer connected FPGA in a GPU server.
%
%The high level diagram of such a test system is shown in Figure~\ref{fig:setup}.
%
%The FPGA is programmed to advertise itself as a large memory just like the host memory.
%
%The FPGA advertises itself as a large memory by using base address register (BAR).
To this end, we add custom logic in the FPGA to monitor the request count, average/peak  number of outstanding memory requests, and request sizes. 
%To this end, we add custom monitoring logic in the FPGA such as a request counter, average/peak  number of outstanding memory requests, and an access size histogram. 
%
%With the help of this monitoring platform, we find that the zero-copy memory region has very similar characteristics to the regular global memory accesses and the same optimization techniques~\cite{cudapractice,cudaoptimize} can be applied to improve the performance.

\subsection{Zero-Copy Mechanism and Optimization}% \fixme{section number to correct}
\label{sec:zerocopy.opt}
%\fixme{may be a bit stupid: should we establish toy example is representative of graph traversal algos?}
%Now that we understand how to enable zero-copy memory, we next need to understand how the zero-copy memory access pattern behaves when accessing external memory using PCIe.
%
Now that we have a way to track zero-copy memory requests, we next need to understand the GPU access pattern to zero-copy memory. %when accessing external memory using PCIe.
We create a toy example where the GPU needs to traverse a large 1D array in a zero-copy region and use a GPU kernel to copy its content to the GPU's global memory.
%To this end, we setup a FPGA to serve as large Zero-copy memory region over PCIe, as described in $\S$~\ref{sec:eval.setup}.
%We create a toy example where the GPU threads copy a large 1D array from this memory region to the GPU's global memory.
%
%\fixme{In this toy example, the large 1D array is representation of CSR edgelists. } \david{David: What? No. This toy example is just a single 1D array. Has nothing to do with CSR. This is a general test. Adopting CSR is explained in Section 4.}
%TODO: NEED TO FIGURE OUT HOW TO PROVIDE INTUTION ON WHY WE NEED TO STUDY MISALIGNED REQUEST here. 
%
The algorithm to solve the toy example can either perform strided access or merged with misaligned access or merged with aligned access. 
%To solve this problem, we write three types of \fixme{codes} where each perform strided access, merged+aligned access, and merged+misaligned access to the 1D array. 
%Vikram does not like the usage of the word codes. Listing perhaps better but still not convinced.
%
%All PCIe traffics generated by the three types of the codes are monitored using the FPGA test platform described.
All PCIe traffic generated by these three variants is monitored using the FPGA monitoring platform and Intel VTune~\cite{Vtune}. %described in $\S$~\ref{sec:eval.setup.fpga}. 
%\fixme{David to review last two sentences}
%\fixme{Vikram: I think in the section 3.2 we should state, LLC is disabled for zero-copy access. The reason is when i reach strided access, it feels like an obvious question to be clarified}
%\david{David: I think figure 2 already shows llc caching is happening}
%With the monitoring platform, 
PCIe layer in Figure~\ref{fig:coalescing} shows the GPU access patterns we observed with the FPGA monitoring platform while trying different CUDA kernels.
We observe that GPU can access the zero-copy memory in four different sizes starting from 32-byte to 128-byte in 32-byte steps.
The access size is dependent on the algorithm access pattern and is described next. 
%We observe the GPU can access zero-copy memory in 4 different accesses sizes, either 32-bytes, 64-bytes, 96-byes, or 128-bytes
%The access size is dependent on the algorithm access pattern described next.
% \david{David: Ok for me}

\textbf{Strided Access:}
In this method, each thread takes a chunk of the 1D-array and iterates over the chunk one element at a time.
%
%A rough access pattern of this method is shown in
This access pattern is illustrated in 
Figure~\ref{fig:coalescing} (a).
With GPU threads iterating over their own neighbor lists, we find that each thread generates a new 32-byte request every time they cross a 32-byte address boundary.
%
%Therefore, if the data type of the array is 4-byte, each request can serve at most 7 more future memory accesses.
Therefore, if the data type of the array is 4-byte, each PCIe request can serve up to 8 memory accesses.
%\fixme{David to review sentence as it is rephrased.}~\david{David: Good to me}
%

%However, this 32-byte request is not even close to being enough to saturate PCIe 3.0 x16 bandwidth for the following reasons.
However, this 32-byte request brings several limitations to the overall system.
First, each PCIe 3.0 transaction layer packet (TLP) has at least an 18-byte of header overhead.
Thus, fetching 32-byte of data makes the PCIe overhead ratio of at least 36\%.
%Even if we assume the PCIe header size is always the smallest, fetching 32-byte data makes the PCIe overhead ratio to 36\%.
%
Second, considering the PCIe latency, the number of outstanding requests to saturate the PCIe interconnect is non-negligible.
With our test platform, we find the PCIe round trip time (RTT) between the GPU and the FPGA is about roughly 1.0us to 1.6us.
%
%When all PCIe request sizes are 32-byte, 1.0us of PCIe RTT requires at least 269 outstanding requests at any given point to reach 8GB/s of effective PCIe bandwidth.
%
By the PCIe 3.0 specification, the maximum number of outstanding requests is 256 as the width of the tag field used to record the outstanding request is 8-bit~\cite{pciespec}.
In this case, the maximum bandwidth we can achieve with only 32-byte requests and 1.0us of RTT is merely 32B / ( 1.0us / 256 ) = 7.63GB/s.
If we assume the PCIe RTT is always 1.6us, the bandwidth decreases to 4.77GB/s.
%However, it is impossible to reach this number of outstanding requests as the GPU can generate at most 256 PCIe requests at any given moment.
%
Third, the minimum memory access size for DDR4 DRAM is 64-byte in the test system.
Considering that DDR4 2400MHz DRAM can provide 19.2GB/s of sequential bandwidth, requesting only 32-byte read requests halves the effective DRAM bandwidth to 9.6GB/s.
Even the overall DRAM bandwidth can be increased by adding more memory channels, this is still very wasteful.
%even if the host DDR4 system offers more bandwidth.
%
Finally, these 32-byte data items will likely occupy GPU cache and can be evicted before all elements are traversed due to cache thrashing.
%accesses of other threads and the IO amplification resulting from moving large pages.
%and make GPU to oversubscribe data. \fixme{<- not sure wats meant by this}
%

\begin{figure}
  \centering
  \includegraphics[width=\linewidth]{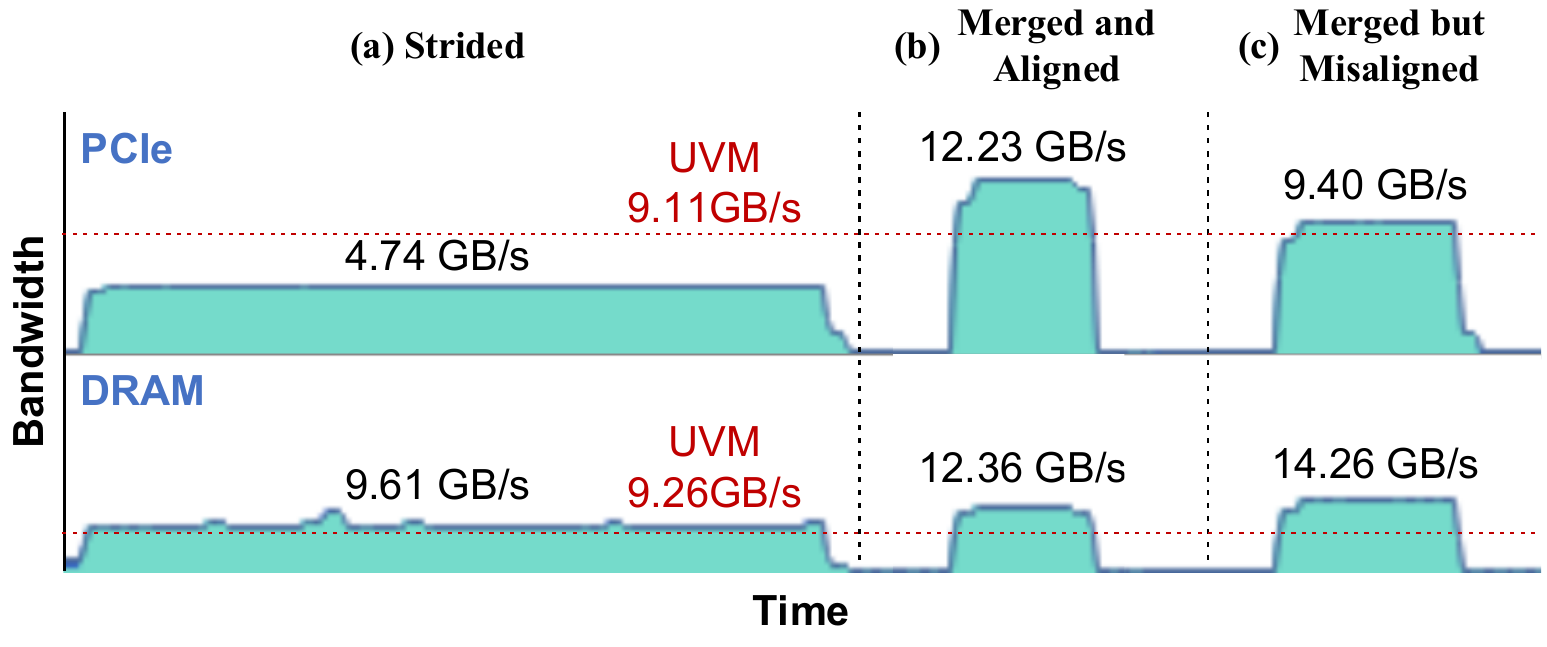}
  \caption{Average PCIe and DRAM bandwidth utilization for the different zero-copy access patterns, as reported by Intel VTune.} 
  %\caption{Runtime average PCIe and DRAM bandwidths of toy example from Intel VTune, using different zero-copy access patterns.}
  \label{fig:vtune}
\end{figure}

Figure~\ref{fig:vtune} shows the average PCIe and DRAM bandwidth utilization over time when executing the traversing kernel as reported by Intel VTune. % profiling tool. %\fixme{Vikram: is the traversing kernel here used is the toy example of real graph?}\david{David: why? I never said a real graph. Only said 1D array.} I thought so. but double checking. 
%Figure~\ref{fig:vtune} shows the plot is directly from Intel VTune~\cite{Vtune} measurement results which show the runtime average PCIe and DRAM bandwidth.
%
The peak bandwidth we achieved with UVM is drawn as a red dashed line in the figure as a reference.
Looking at Figure~\ref{fig:vtune} (a), we can clearly identify the limitations previously described. 
%Figure~\ref{fig:vtune} (a) shows the limitations \fixme{in practice clearly}. %rephrase needed. 
%
The amount of data that needs to be read from DRAM is doubled to serve 32-byte PCIe requests.
%As explained above, the amount of data needs to be read from DRAM is doubled to serve 32-byte PCIe request, and therefore the achieved bandwidth is also about double of the PCIe bandwidth.
%
The PCIe bandwidth is also far from the maximum PCIe 3.0 x16 bandwidth as the number of outstanding requests is not enough and the per-request PCIe overhead is significant.
Furthermore, it results in transferring more bytes to the GPU compared to the original dataset size due to the frequent cacheline evictions.
%
%The key to address these limitations is by performing aligned merged accesses.
To address these limitations the key is to align and merge accesses.
%as in the case of global memory access \fixme{is it necessary to mention the case of global memory access here, seems out of no where}. 
%
%Next, we will analyse what happens to the average PCIe and DRAM utilization when we use merged and aligned accesses. 
We analyze the PCIe and DRAM bandwidth utilization with these optimizations next.

%\fixme{Vikram: TO ADD A PARAGRAPH AS MOTIVATION FOR NEXT TWO EXPERIMENTS. For Strided case, we have 36\% overhead which means max achievable PCIe BW is ~10GBps. However we get about 4.74GBps. To quantify what is important - between the coalescing vs alignment - we perform next two experiments. Merged + aligned vs merged + unaligned. }
%\david{David: Not sure what you mean by it should be 10GB/s. I said there are multiple limitations. Why only saying 36\% PCIe overhead?}

\textbf{Merged and Aligned Access:}
In this case, threads are grouped into warps, with each warp containing 32 threads,
%all threads are merged in a warp granularity (32-threads) 
%and all threads in a warp access the input array a row at the same time.
and the threads in a warp access consecutive elements in a 128-byte cacheline of the input array.
%at a time.
%
This allows the GPU coalescing unit to automatically merge the contiguous 32-byte memory requests into a single larger 128-byte PCIe request (Figure~\ref{fig:coalescing} (b)). 
%This access pattern is similar to when accessing global memory.
%
With 128-byte PCIe requests, it becomes much easier to reach the maximum PCIe bandwidth.
First, the PCIe TLP overhead ratio decreases from 36\% to 12.3\% .
Second, having only 135 PCIe outstanding requests is sufficient to reach 16GB/s of bandwidth (without considering other PCIe overheads).
Lastly, 128-byte is a multiple DRAM request size and therefore there is no inefficiency in the DRAM interface.
In Figure~\ref{fig:vtune} (b), we see this approach can saturate the PCIe bandwidth at about 12.23GB/s
, matching the measured bandwidth when using the \texttt{cudaMemcpy()} API to perform a block data transfer.
%while the maximum bandwidth we measured with \texttt{cudaMemcpy()} i.e., about 12GB/s.
%
%Moreover, thanks to the aligned accesses, there is nearly no wasted DRAM bandwidth to serve the PCIe memory requests.

\textbf{Merged but Misaligned Access:} 
%\fixme{I would reorder aligned and unaligned access para. First speak about unaligned and then aligned.} \david{David: For now, I'll keep this ordering.}
%
%\fixme{Connecting sentence. }
%
However, for all practical purposes, guaranteeing 128-byte alignment 
%on a \fixme{CSR data structure}\david{David: Let's not be too specific where this is going here. We will talk about the CSR misalignment in Section 4. This section is supposed to be just very generic, not targeting anything.}
for any data structure can be difficult. 
It is possible that the starting index of a warp is not aligned with the 128-byte boundary.
%
%However, in more realistic cases, it is possible that the starting index of each warp is not aligned with the 128-byte boundary.
%
Some warps may need to make two separate PCIe requests to fetch a single 128-byte cacheline.
%
%In the worst case, if all warps are accessing memory side-by-side and the first warp's memory access is not 128-byte aligned, the misalignment is cascaded to all subsequent warps.
In the worst case, if a warp's memory access is not 128-byte aligned and warps access contiguous regions of memory, the misalignment can be cascaded to all subsequent warps.
Unfortunately, this results in all warps generating two PCIe requests.
%Unfortunately, in this case, every warp needs to generate two PCIe requests.
%
In Figure~\ref{fig:coalescing} (c), we show an emulated misaligned case where each warp is intentionally accessing memory offset by 32-byte from 128-byte boundary and therefore all warps end up generating a 32-byte and a 96-byte PCIe request.
From Figure~\ref{fig:vtune} (c), we can see the achieved PCIe bandwidth is lower than the aligned case.
To avoid this, either the starting index of warps should be shifted or the input data must be shifted in memory so the data accessed first is 128-byte aligned.
%should be aligned to 128-byte boundary from the beginning.

\section{EMOGI: Zero-Copy Graph Traversal}
\label{sec:implementation}
Now that we understand zero-copy memory and its characteristics, we discuss how to efficiently use zero-copy memory for graph traversal when the graph cannot fit in the GPU memory. 
%In this section, we explain the application of zero-copy in traversing graphs which do not fit in GPU memory.
%
First, we describe the micro data locality we observed in graph traversal applications to justify why zero-copy should perform better than UVM (see $\S$~\ref{sec.emogi.datalocal}).
Then, we introduce our baseline graph traversal algorithm  (see $\S$~\ref{sec.emogi.baseline}) and optimize it for zero-copy memory based on the knowledge we gathered from $\S$~\ref{sec:zerocopy.opt} (see $\S$~\ref{sec.emogi.opt}).

\subsection{Data locality in Graph Traversal}
\label{sec.emogi.datalocal}
To exploit zero-copy for graph traversal, we preferably need at least 128-byte of spatial locality to best use each memory access.
A single 128-byte zero-copy access can have 16 or 32 elements of data if the CSR data type is 8-byte or 4-byte, respectively.
%Using a 4-byte data type, 128-byte zero-copy memory requests can have 32 elements of data while using a 8-byte data type, it can hold 16 elements per zero-copy memory request. 
%
%\fixme{When we assume the input data type size is 4-byte, it is identical to 32 elements of data.
%For 8-byte data type, it is only about 16 elements of data.}
%\fixme{David are you just saying with 4B we can have 32 elements while 8B we can have 16 elements in 128B request? I dont understand the identical part here}\david{David: Your understanding is correct}
%
Compared to UVM, which requires at least 4KB of spatial locality (512 or 1024 elements of data), finding 16 to 32 elements of spatial locality is reasonable for the graphs we studied. %Vikram: Obvious comment - This is entirely dependent and cant be generalized. We should tell here for the graphs we studied this is easy to get.}. 

Based on our analysis of 1122 graphs 
%that has at least 1M of vertices and edges 
from Network Repository~\cite{networkrepo}, SuiteSparse Matrix Collection~\cite{suitesparse}, and  LAW~\cite{ubicrawler}, we find the average degree per vertex is 71.
This means, when those graphs are represented in an adjacency list format like CSR, each vertex's neighbor list is 71 elements long on average.
%
%\fixme{INFO:To make this section even stronger we should so 90\% of node have average degree of more than 16 elements over just average. If time permits, we can add a graph too.}\david{David: Based on my experiment, not 90\% of nodes have more than 16. Apparently the average matters more}
%
Considering that graph traversal algorithms require scanning the entire neighbor list of a vertex, we can obtain a spatial locality of 71 elements on average in graphs. %assume in average there are 71.27 elements long spatial locality in graphs.
Such a spatial locality can benefit from efficient 128-byte requests to zero-copy memory.
%Since zero-copy only needs 16 to 32 elements to exploit data locality in each request, such a large average spatial locality in graph can benefit from efficient 128-byte requests. 
In contrast, it is more difficult to achieve the same level of efficiency using UVM since the available spatial locality is significantly less than the required 512 or 1024 elements.
%over in-efficient 32-byte requests. \fixme{i dont think we need that last part}
%Even if we consider some variations between vertices and graphs, this is sufficiently large enough to generate more 128-byte requests than 32-byte requests.
\subsection{\pname{} Baseline}
\label{sec.emogi.baseline}

\pname{} assumes the input graph is stored in the memory using the CSR data layout (see $\S$~\ref{subsec:par-graphtraversal}). 
All input data structures are statically mapped during initialization.
%
%The edge list that does not fit in GPU memory is allocated in the host memory, but other small data structures such as vertex list and algorithm specific temporary structures are allocated in GPU memory. 
The edge list is allocated in the host memory as it doesn't fit in GPU memory, but other small data structures such as buffers and the vertex list are allocated in GPU memory.
It is worth noting that even for the biggest graphs we evaluated (see  $\S$~\ref{sec:eval.method}), the vertex list consumes at most 1GB of memory while the edge list can consume 38GB.
% {\color{red} WMH: Let's double check the 1GB number. It seems to be small in the sense that our biggest graphs have only less than 256M vertices?}
%Vikram: Assuming 4B, 256M vertices is about 1GB. Our graphs used in evaluation are lesser than 256M vertices. We are considering to a submission now. We have about 12 more mins to the deadline.
Thus, GPU memory is sufficient for the vertex list.

\pname{} adopts vertex-centric graph traversal algorithms. 
For every vertex that needs to be processed, a worker is assigned and the worker traverses a neighbor list associated with the vertex in the edge list.
Listing~\ref{lst:naive} shows the pseudo code of our naive baseline implementation.
Here, the worker is a single GPU thread and each worker is assigned to the neighbor list associated with its corresponding vertex.
When each neighbor list is larger than 128-byte, this baseline implementation has a similar memory access pattern to the strided case explained in $\S$~\ref{sec:zerocopy.opt}.
%
%\fixme{vikram here}

Compared with the UVM approach, \pname{}'s graph traversal approach removes the page faults from occurring and reduces the I/O amplification as only the needed bytes are moved.  
%From this implementation, the only advantages we can expect over the UVM approach are a removal of page faults and potentially smaller data oversubscription.
%
In the vertex-centric graph traversal approach, the input graph is traversed by a single vertex depth on every kernel execution. 
%For every each kernel execution, the input graph is traversed by single vertex depth.
%
Therefore the total number of kernels launched, say in the case of breadth-first-search (BFS), is equal to the distance between the source vertex to the furthest reachable vertex.

\subsection{Optimizations}
\label{sec.emogi.opt}
%Since the \pname{} baseline implementation is similar to the strided case presented in $\S$~\ref{sec:zerocopy.opt}, it comes with a draw-back: uncoalesced memory requests. 
Since the \pname{} baseline implementation is similar to the strided case presented in $\S$~\ref{sec:zerocopy.opt}, it suffers from uncoalesced memory requests. 
As we noted, without addressing this, one cannot generate efficient PCIe requests to the zero-copy memory. 
In this subsection, we will discuss how \pname{} addresses this limitation using the insights from $\S$~\ref{sec:zerocopy.opt} and modifying only the GPU kernel code of the traversal application.
Thus, it is entirely possible to package the proposed optimizations into a library to lessen the programmer's effort when trying to exploit them.

\subsubsection{Merged Memory Access:} %Tile of the algorithm and this subsection should be same.
\label{sec.emogi.mam}
\pname{} performs merged memory accesses in per vertex granularity, similar to ~\cite{maxwarp}. 
The difference between \pname{} and ~\cite{maxwarp} is that \pname{} always fixes the worker size to an entire warp (i.e., 32 threads).
Thus a whole warp is responsible for traversing the neighbor list of one vertex.
The specific implementation of this optimization is explained with red comments in Listing~\ref{lst:aligned}.
%We note that this optimization only requires changes to the GPU kernel for traversing graphs and launching the kernel with more threads.
%One small difference from ~\cite{maxwarp} is that we always fix the worker size to an entire warp granularity (32-threads).
%
This allows \pname{} to always optimize for generating the maximum sized PCIe request to the zero-copy memory. 
%This is because we always look for the opportunity to generate the maximum sized PCIe requests.
%
If the input graph fits in the GPU memory and the average degree of vertices in the graph is small, fine-tuning the worker size 
%to smaller than the warp granularity 
could potentially reduce the number of idle threads during each fetch, exploit more memory parallelism, and ultimately utilize GPU global memory bandwidth more efficiently.
%When input graphs fit in GPU memory and the ratios of edge to vertex are small in average, fine-tuning the worker size to something smaller than 32-thread can potentially generate more memory requests and further utilize the global memory bandwidth.
%
However, \pname{}'s primary goal is to achieve good performance on graphs that do not fit in the GPU memory and it requires fetching data over an external interconnect that is about 10-100$\times$ slower than the GPU global memory. 
In this case, fine-tuning and reducing the worker size cannot add any additional benefit as there is no further room to accept more memory requests in the already constrained interconnect. 
%However, in our case, the external interconnect is about 10 to 100 times slower than the global memory and usually there is no more room to accept more memory requests.
%
In fact, making smaller memory requests can have an adverse effect and decrease the effective bandwidth.
%in opposite.  
%
Empirically we observed when the interconnect bandwidth is low, a large number of threads are idle. Therefore, assigning a 32-thread warp to fetch data for even vertices with very few neighbors results in acceptable performance.  
%In general, there are always large number of idling threads due to the low interconnect bandwidth and therefore assigning 32-thread warp for very low degree vertices as well is acceptable.

%\begin{lstlisting}[float,caption={Coalesced Memory Access},label={lst:coalesced}]
%#define WARP_SIZE 32
%
%void coalesced(*edgeList, *offset, ...) {
%    thread_id = get_thread_id();
%    lane_id = thread_id % WARP_SIZE;
%    warp_id = thread_id / WARP_SIZE;
%    ...
%    
%    start = offset[warp_id];
%    end = offset[warp_id + 1];
%    
%    <@\textcolor{blue}{// Multiple threads access a single input chunk}@>
%    <@\textcolor{blue}{// at the same time in a row}@>
%    for (i = start; i < end; i += WARP_SIZE) {
%        edgeDst = edgeList[i + lane_id];
%        ...
%    }
%    ...
%}
%\end{lstlisting}

\begin{lstlisting}[float,caption={Uncoalesced Memory Access},label={lst:naive}]
void strided(*edgeList, *offset, ...) {
    thread_id = get_thread_id();
    ...
    start = offset[thread_id];
    end = offset[thread_id + 1];
    
    <@\textcolor{blue}{// Each thread loops over a chunk of edge list}@>
    for (i = start; i < end; i++) {
        edgeDst = edgeList[i]; ...
    } ...
}
\end{lstlisting}

\begin{lstlisting}[float,caption={Coalesced Memory Access (Merged + Aligned)},label={lst:aligned}]
#define WARP_SIZE 32

void aligned(*edgeList, *offset, ...) {
    thread_id = get_thread_id();
    lane_id = thread_id % WARP_SIZE;
    <@\textcolor{red}{// Group by warp}@>
    warp_id = thread_id / WARP_SIZE;
    ...
    start_org = offset[warp_id];
    <@\textcolor{blue}{// Align starting index to 128-byte boundary}@>
    start = start_org & ~0xF; <@\textcolor{blue}{// 8-byte data type}@>
    end = offset[warp_id + 1];
    
    <@\textcolor{red}{// Every thread in a warp goes to the same edgelist}@>
    for (i = start; i < end; i += WARP_SIZE) {
        <@\textcolor{blue}{// Prevent underflowed accesses}@>
        if (i >= start_org) {
            edgeDst = edgeList[i + lane_id];
            ...
        }
    } ...
}
\end{lstlisting}

\subsubsection{Aligned Memory Access}
\label{sec.emogi.maa}
%\fixme{vikram here. }
As we discussed in $\S$~\ref{sec:zerocopy.opt}, a misaligned access to the 1D data array can result in 
%generating 
multiple smaller zero-copy requests.
% over the external interconnect. 
%Even if the memory requests are merged, it can still cause performance degradation (see Figure~\ref{fig:vtune}(c)).
To address this, we have to not only merge memory accesses but align them as well.
%perform aligned access along with the merged memory access. 
However, doing this on a CSR edge list is not straightforward. 
This is because CSR doesn't align the edge list as alignment requires padding and thus increases memory footprint.
%For the maximum data compression rate, data alignment is not required for CSR.
%
Starting addresses of neighbor lists for graphs stored in CSR can be at any location in the memory.

One way to address this challenge is to pre-process the CSR graphs and align neighbor lists to 128-byte boundaries.
However, this might incur excessive memory overhead.
More importantly, one of the goals of this work is to avoid any pre-processing.

Therefore, instead of manipulating the input data, we force all warps to 
%be shifted to a
start from the closest preceding  128-byte boundary when there is misalignment.
For instance, as shown in Listing~\ref{lst:aligned} with blue comments, all starting indices fetched from the offset array is shifted to the closest 128-byte boundary before the list.
With this change to the GPU kernel code, all subsequent warp memory accesses are guaranteed to have 128-byte alignment.
Of course, some of the threads in the warp must be turned off during the first iteration of data fetching with a conditional statement to prevent reading unnecessary bytes.
%so that no memory accesses to the locations before the actual starting point would occur.
%
Similar to the memory access merge optimization, this additional conditional statement increases the occurrence of control divergence in CUDA kernels. 
However, due to the high external interconnect latency, it is more important to not miss any opportunity for generating large memory requests.
%
%Furthermore, even though the optimizations require changes to GPU kernel code for graph traversal, it is entirely possible to package these changes into a library to lessen the programmer's effort when trying to exploit these optimizations.
%\fixme{Lets see if really need an additional experiment or we just need some profiling data and we can write it out. What we need to show is increasing control divergence from X to X+delta is still better. We can add this data in 5.3.3 to justify it is fine.}
%
%Our empirical evaluation showed the cost of increased control divergence is smaller than performing a misaligned access to zero-copy memory.  {\color{red} Would it be sufficient just to state that we are limited by interconnect bandwidth even with the extra control divergence so the control divergence does not cost any execution time?}

%\fixme{Since we prove that the application is limited by IO bandwidth for most graphs in the evaluation, control divergence should not matter. However, we might still have to state how much increase in the control divergence are we speaking off on average across all the graphs. }

\section{Evaluation}
\label{sec:eval}
Our evaluation shows that 
(1) \pname{} improves the performance of graph traversal algorithms by efficiently accessing the zero-copy memory for very large graphs,
(2) \pname{} is mainly limited by the PCIe bandwidth and it scales almost perfectly linearly when PCIe 3.0 is replaced with PCIe 4.0,
(3) \pname{} remains performant even with the latest generation of GPU NVIDIA Ampere A100~\cite{amperewhite} and achieves better scaling compared to the UVM optimized implementation.
%Before we dive into each of these evaluation, we describe our experimental setup. 

\subsection{Experiment Setup}
\label{sec:eval.setup}

\subsubsection{System Overview:}
We use a Cascade-lake server machine with two 20 core Intel Xeon Gold 6230 CPUs equipped with 256GB of DDR4 2933MHz memory and an NVIDIA Tesla SXM2 V100 16GB GPU as our evaluation platform.
The system is configured as shown in Figure~\ref{fig:setup}. 
We use the FPGA only to analyze the zero-copy memory access pattern across different graphs.
The detailed system specification is provided in Table~\ref{tab:sysconfig}. 
Graph edge lists are stored in the host memory while the vertex list and other temporary data structures are stored in the GPU memory.

\subsubsection{Systems Compared:}
% \subsection{Systems Compared:}
% \label{sec:eval.setup}

% \fixme{Table~\ref{tab:sysconfig}}
To show the performance benefit of \pname{}, we use four different graph traversal algorithms: Breadth-First Search (BFS), Single-Source Shortest Path (SSSP), Connected Components (CC), and PageRank (PR)~\cite{ilprints422}. 
We base our initial implementation of BFS and SSSP from ~\cite{graphbig, topowarp}, CC baseline implementation from ~\cite{gardenia}, and PR baseline implementation from ~\cite{galois}. %We add the needed modifications to support large graphs, UVM and any required optimization. 
%\fixme{add a sentence make it extended to other graph traversal algo.}
We compare \pname{} with the following systems: 

\textbf{(a) UVM} implementation stores the CSR edge list in the UVM address space while the vertex list is kept in the GPU memory. In addition, the CSR edge list in the UVM address space is marked as \texttt{cudaMemAdviseSetReadMostly} using the \texttt{cudaMemAdvise()} CUDA API call. This optimization allows the GPU to create a read-only copy of the accessed pages in the GPU's memory.
We also tested other available UVM driver flags but did not observe notably differences.
%We also tested other available flags and UVM hints and we found the performance hardly differs when compared to using only ~\texttt{cudaMemAdviseSetReadMostly}.
\textbf{(b) Naive} implementation is the baseline implementation of \pname{} using zero-copy memory and is identical to Algorithm~\ref{lst:naive}. In this implementation, the vertex list is stored in the GPU memory while the edge list is kept in the zero-copy host memory.
\textbf{(c) Merged} implementation of \pname{} merges the memory requests to the zero-copy memory, as discussed in $\S$~\ref{sec.emogi.mam}. However, in this implementation, there is no guarantee that accesses to the zero-copy memory are aligned.
\textbf{(d) Merged+Aligned} implementation is the fully optimized version of \pname{} where the memory accesses are not only merged but we force all warps to shift to the 128-byte boundary when there is a misalignment. This implementation is discussed in $\S$~\ref{sec.emogi.maa}.

% We use a Cascade-lake server machine with two 20 core Intel Xeon Gold 6230 CPUs equipped with 256GB of DDR4 2933MHz memory and a NVIDIA Tesla SXM2 V100 16GB GPU as our evaluation platform.
% %
% The system is configured as shown in Figure~\ref{fig:setup}. The detailed system specification is provided in Table~\ref{tab:sysconfig}.

\begin{table}[t]
  \caption{Evaluation system configuration.}
  \label{tab:sysconfig}
  \begin{tabular}{cl}
    \toprule
        Category & \multicolumn{1}{c}{Specification} \\
    \midrule
        CPU & Dual Socket Intel Xeon Gold 6230 20C/40T \\
    \midrule
        Memory & DDR4 2933MHz 256GB in Quad Channel Mode\\
    \midrule
        GPU & Tesla V100 HBM2 16GB, 5120 CUDA cores \\
    \midrule
        OS   & CentOS 8.1.1911 \& Linux kernel 5.5.13  \\
    \midrule 
        S/W & NVIDIA Driver 440.82 \& CUDA 10.2.89\\
    \bottomrule
  \end{tabular}
\end{table}

\subsection{Evaluation Datasets}
\label{sec:eval.method}
For the evaluation, we use the graphs listed in Table \ref{tab:dataset}. GK, GU, FS, and ML are the largest four graphs from SuiteSparse Matrix Collection~\cite{suitesparse} and SK, and UK5 are commonly used large graphs from LAW~\cite{ubicrawler}.
This collection of graphs covers data from different areas such as biomedicine, social networks, web crawls, and even synthetic graphs.
All the graphs, except for SK and UK5, are undirected.
We use the default weights for GU, GK, and ML graphs while we randomly initialize weights for the rest of the graph from the integer values between 8 to 72. 
%Weights are represented in 4-byte datatype.
%
The average degree of the graphs is 38, except for the ML graph, which has an average degree of 222.
For fair BFS and SSSP performance evaluations, we pick 64 random vertices from each graph as the starting sources and reuse the selected vertices for all measurements.
The final execution time is calculated by averaging the execution times of the 64 cases, but some results are removed from the average when the selected vertices have no outgoing edges.
Edge weight values are only used by the SSSP algorithm.
%
%For a better understanding of the distribution of the neighbor list sizes of each graph, we also draw a number of edges cumulative distribution function (CDF) for each evaluation graph as shown in Figure~\ref{fig:graph_dist}.
% For a better understanding of the distribution of the neighbor list sizes in the graph, we also plot the cumulative distribution function (CDF) on the number of edges in each graph, as shown in Figure~\ref{fig:graph_dist}.
%
% The horizontal axis of this CDF is cut to 96 since many of the graphs have vertices with extremely large number of degrees.
% %
% Using the CDF, we identify that the total number of connecting vertices with less than 16 degrees is less than 10\% of the entire edges. \fixme{ok, so wats the point?}
%
%We will discuss about this figure in more detail in $\S$~\ref{sec:eval.memcoal}.

\begin{table}[t]
  \caption{Graph Datasets. $V$ = Vertex, $E$ = Edge, and $w$ = Weight.}
  \label{tab:dataset}
  \begin{tabular}{@{\extracolsep{4pt}}clcccc@{}}
    \toprule
    \multirow{2}{*}{Sym.}  & \multicolumn{1}{c}{\multirow{2}{*}{Graph}} & \multicolumn{2}{c}{Number} & \multicolumn{2}{c}{Size (GB)} \\
    \cline{3-4}
    \cline{5-6}
            &                                             & $|V|$    & $|E|$ & $|E|$& $|w|$ \\
    \midrule
    GK      & GAP-kron        \cite{GAP}                  & 134.2M   & 4.22B & 31.5 & 15.7 \\
    GU      & GAP-urand       \cite{GAP}                  & 134.2M   & 4.29B & 32.0 & 16.0 \\
    FS      & Friendster      \cite{Friendster}           & 65.6M    & 3.61B & 26.9 & 13.5 \\
    ML      & MOLIERE\_2016   \cite{MOLIERE_2016}         & 30.2M    & 6.67B & 49.7 & 24.8 \\    
    SK      & sk-2005         \cite{BoVWFI,BRSLLP,BCSU3}  & 50.6M    & 1.95B & 14.5 & 7.3
    \\
    UK5     & uk-2007-05      \cite{BoVWFI,BRSLLP}        & 105.9M   & 3.74B & 27.8 & 13.9 \\
    \bottomrule
  \end{tabular}
\end{table}

\subsection{Case-Study: Breadth-First Search}

In this section, we take BFS as an example and thoroughly evaluate PCIe traffic for request size distribution, achieved bandwidth, and the total amount of data transferred.
%
%In addition, to understand the performance scalability of \pname{} with higher bandwidth interconnects, we also measure the BFS total execution time using NVLink.
%
Throughout the evaluation, we use the UVM implementation as the baseline.
% %
% \fixme{Add where the source code is taken from.}

% \fixme{speak about the performance difference between naive, merge and merge+aligned here for different graphs. }

\begin{figure}[t]
  \centering
  \includegraphics[width=\linewidth]{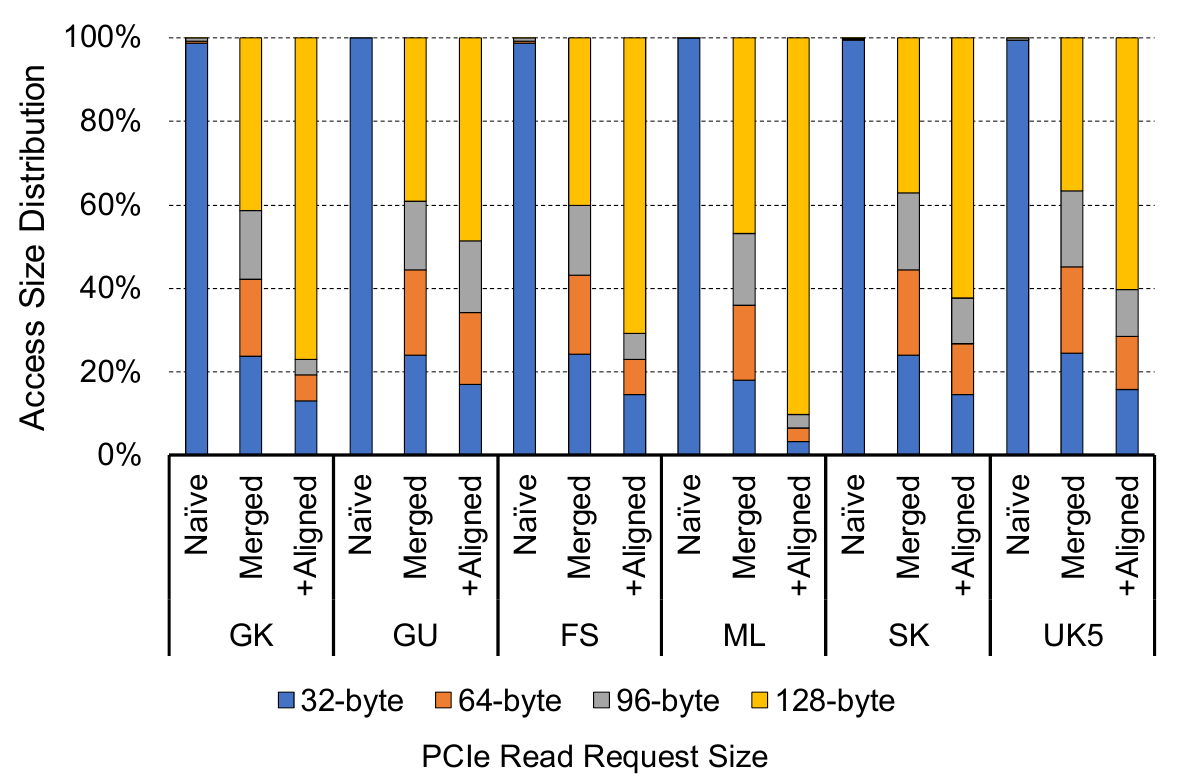}
  \caption{Distribution of PCIe read request sizes in BFS. \texttt{+Aligned} is abbreviation for \texttt{Merged+Aligned}. As the merged and aligned optimizations are added, the BFS application generates more 128-byte requests for efficient access.}
  %\caption{Distribution of number of PCIe read requests with different access sizes in BFS. \texttt{+Aligned} is abbreviated for the \texttt{Merged+Aligned} implementation. These results are collected using FPGA.}
  \label{fig:request_size}
\end{figure}

\begin{figure}[t]
  \centering
  \includegraphics[width=\linewidth]{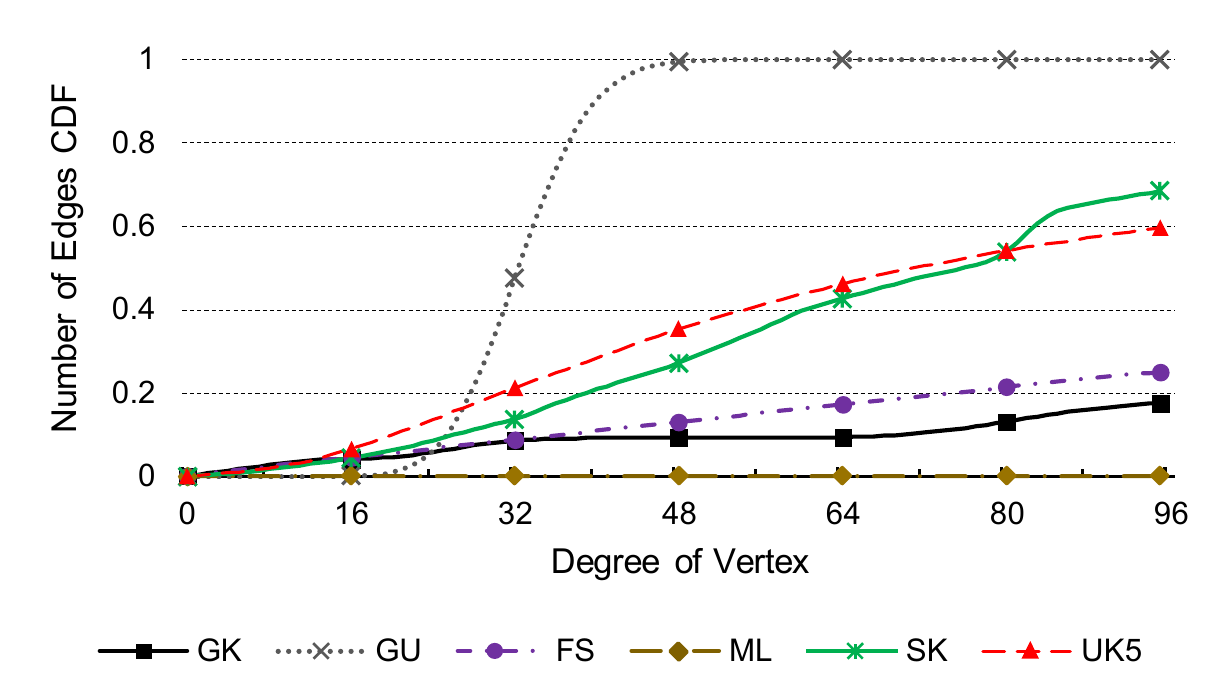}
  %\caption{Number of edges CDF of evaluation graphs. This plot provides us a better understanding of the distribution of the neighbor list sizes in the graphs.}
  \caption{Number of edges CDF of evaluation graph. This plot provides us a better understanding of the distribution of the neighbor list sizes in the graphs. For example, the GU graph has all of its edges associated with vertices with degree between 16 and 48, meaning the neighbor lists contain at most 48 neighbors.}
  \label{fig:graph_dist}
\end{figure}

\begin{figure}[t]
  \centering
  \includegraphics[width=\linewidth]{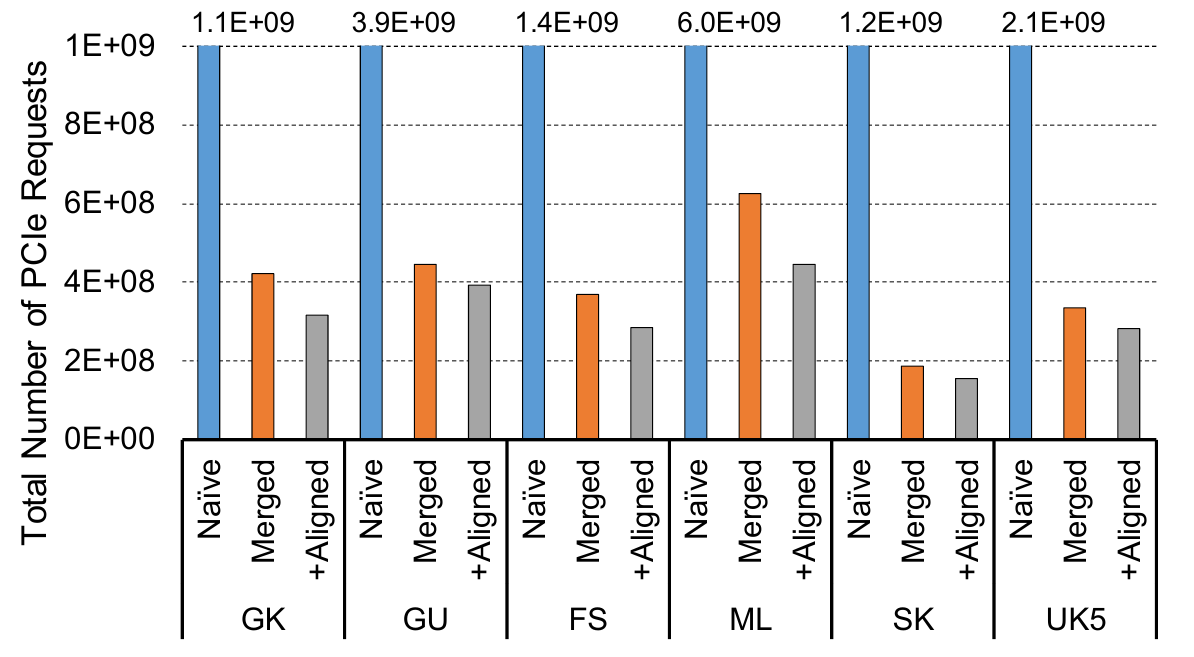}
  \caption{Number of PCIe requests sent for \texttt{Naive}, \texttt{Merged} and \texttt{Merged+Aligned} implementations while executing BFS on various graph. Collected from FPGA. \texttt{Merged} optimization reduces the PCIe memory requests by up to 83.3\% compared to the \texttt{Naive} implementation. \texttt{Merged+Aligned} optimization can further reduce the PCIe memory requests by up to 28.8\%. %compared to the \texttt{Merged} implementation.
  \texttt{+Aligned} is abbreviation for \texttt{Merged+Aligned}.}   
  \label{fig:reqsizedetail}
\end{figure}

\subsubsection{Zero-copy Request Size Distribution:}
\label{sec:eval.memcoal}
In this evaluation, we show the impact of optimizing the memory access pattern from $\S$~\ref{sec:zerocopy.opt} on generating different sizes of PCIe request.
The histogram of the PCIe request size is gathered using the FPGA monitoring platform explained in $\S$~\ref{sec:eval.setup.fpga}.
In Figure~\ref{fig:request_size}, we show the breakdown of request sizes for all the PCIe requests from the three implementations: \texttt{Naive}, \texttt{Merged}, and \texttt{Merged+Aligned}. 
%
%\fixme{We do not measure the UVM case as the UVM direct memory access (DMA) is always done in at least 4KB granularity. is this relevant?}\david{David: Just in case someone curious why there is no distribution of UVM. If it's too obvious then no need to write.}
% it seems redundent as we are renaming it as zero-copy optomization. 
% The \texttt{+Aligned} method includes the both the merge optimization (i.e., \texttt{Merged+Aligned}) \fixme{this is so convoluted.} 
%

We observe in Figure~\ref{fig:request_size} that nearly all PCIe requests in the case of \texttt{Naive} implementation are of 32-byte granularity.
This is because it is only possible to generate a PCIe request larger than 32-byte in the \texttt{Naive} implementation when multiple neighbor lists happen to be spatially near in the edge list and they are accessed by multiple threads in a single warp.
However, such a scenario is extremely unlikely. % and as we can observe from the result nearly all PCIe requests are in 32-byte granularity.
For example, we observe that only 1.3\% of the PCIe requests from BFS on the FS graph are of a size bigger than 32-bytes.
%For the FS graph, we do observe 64, 96 and 128-byte request sizes and the sum of all those three requests is about 1.3\% of the total PCIe requests.
%The one with the highest number of 64, 96, and 128-byte request is FS graph where the sum of those three requests is about 1.3\% of the total PCIe requests.
%

%Moreover, when we analyze the request size distribution for the \texttt{Merged} and \texttt{Merged+Aligned} optimized implementations, we observe 2 oddities.
When we analyze the request size distribution for the \texttt{Merged} and \texttt{Merged+Aligned} optimized implementations, we observe the following.
First, although with the \texttt{Merged} approach the percent of 128-byte requests increases to about 40\% on average, the percent of 128-byte requests is slightly higher than average for the ML graph, at about 46.7\%.
Second, when using the \texttt{+Aligned} approach on graphs that have most of their edges associated with high-degree vertices, we expect that most zero-copy memory requests should be for 128-bytes. 
This is expected because in the \texttt{+Aligned} implementation, zero-copy memory requests are merged and aligned to 128-byte granularity whenever possible.
We observe this behavior for most graphs in Figure~\ref{fig:request_size}.
For example, the percent of 128-byte requests improves by 1.86$\times$ for the GK graph between the \texttt{Merged} and \texttt{+Aligned} implementations.
However, the percent of 128-byte requests improves by only 1.25$\times$ between the two implementations on the GU graph, a graph that has a similar number of edges and vertices as the GK graph.

To further analyze these behaviors, we plot in Figure~\ref{fig:graph_dist}, the cumulative distribution function (CDF) on the number of edges in each graph. 
CDF on the number of edges provides us a better understanding of the distribution of the neighbor list sizes in the graph. 
The horizontal axis of this CDF is cut to 96 as many of the graphs have vertices with an extremely high degree.
%
% Using the CDF, we identify that the total number of connecting vertices with less than 16 degrees is less than 10\% of the entire edges.
From Figure~\ref{fig:graph_dist}, we see that the ML graph has nearly no edges associated with small degree vertices.
Thus, with the \texttt{Merge} optimization many requests can be merged to 128-bytes for the ML graph.
%This explains why the \texttt{Merge} optimization has more 128-byte requests for the ML graph when compared to rest of the graphs. 
The other graphs, like FS, have some edges associated with small degree vertices. Thus not all of their requests can be merged.
%On the other hand, for FS, SK and UK5 graphs we observe some edges associated with lower degree vertices, thus explaining why we cannot make all the zero-copy accesses 128-bytes.
%With the addition of \texttt{+Aligned} optimization, zero-copy memory is now able to send more efficient 128-byte requests. \zaid{<-not understanding this sentence, sounds very generic, wat does the cdf graph tell us that helps us understand this?}
Due to the fact that most vertices have long neighbor lists in the ML graph, 
the \texttt{+Aligned} optimization further maximizes the 128-byte zero-copy accesses, as shown in Figure~\ref{fig:request_size}, and, as a result, reduces the total number of zero-copy memory requests by 28.8\%, as shown in Figure~\ref{fig:reqsizedetail}.
%For the ML graph, the \texttt{+Aligned} optimization further maximizes the zero-copy access sizes to 128-byte access granularity and also reduces the total number of zero-copy memory request by 28.8\%, as shown in Figure~\ref{fig:reqsizedetail}. 
%Thus, the \texttt{+Aligned} optimization provides both bandwidth improvements and also assists in reducing I/O amplification. 
%

% is slight higher than the others due to very high average degree of vertex.
%
% From Figure~\ref{fig:graph_dist}, the graph that has the most of its edges associated with the high degree vertices is the ML graph. 
% As we can note from Figure~\ref{fig:request_size}, most of the requests to the zero-copy memory is now at 128-byte granularity. 
% we observe direct relationship between the actual PCIe request size distribution and CDF on the number of edges.
% %
% From Figure~\ref{fig:graph_dist}, we can see ML has nearly no edges associated with small degree vertices.
% %
% Therefore, the number of 128-byte PCIe requests is overwhelming other PCIe requests.
%

To understand why the request size distribution of GK and GU graphs are significantly different for the \texttt{+Aligned} optimization, we need to understand the neighbor list size distributions of these graphs.
The neighbor lists of the GK graph are extremely unbalanced while the GU graph has uniformly low degrees varying from 16 to 48. 
%To understand the request size distribution of GU, we need to understand few other things first.
%
If we assume the starting location of each neighbor list is uniformly random, then the chance of each neighbor list starting at the exact 128-byte boundary is only 6.25\% when the data type size is 8-bytes.
Therefore, in most cases, the neighbor lists of graphs are not aligned at the 128-byte boundary by default.
If the neighbor list sizes are extremely unbalanced, like in GK, then the misalignment is less problematic since the vertices with high degrees can amortize the cost of the one-time misalignment fix.
%quickly with the following accesses.
%
However, if all vertices have uniformly low degrees, like in GU, then there is no opportunity to amortize the cost of the one-time misalignment fix per vertex.
Due to this, among all the graphs evaluated, only GU shows very little improvement with the \texttt{+Aligned} optimization.

\subsubsection{PCIe Bandwidth Analysis:}
\label{sec:pcieband}
\begin{figure}[t]
  \centering
  \includegraphics[width=\linewidth]{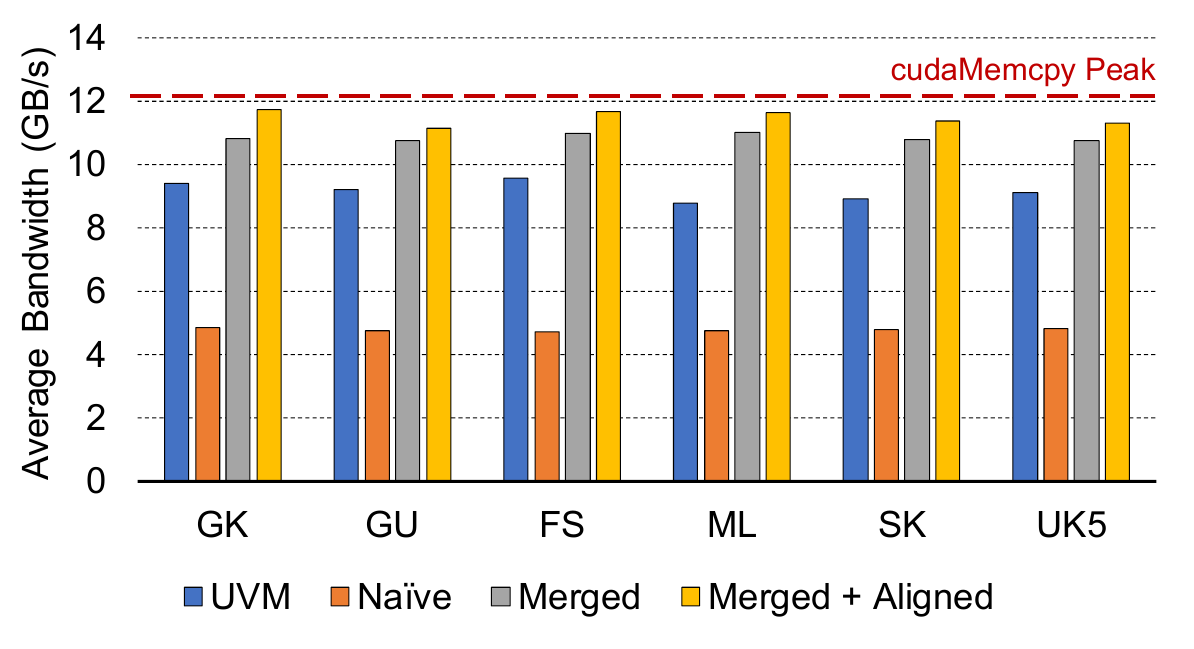}
  \vspace{-20pt}
  \caption{Average PCIe 3.0 x16 bandwidth utilization of the different implementations executing BFS.}
  \vspace{-10pt}
  %\caption{BFS runtime average bandwidth comparison based on PCIe 3.0 x16 interconnect using different methods. \texttt{Merged+Aligned} implementation is nearly able to saturate the available PCIe bandwidth.}
  \label{fig:bandwidth}
\end{figure}

The bandwidths we measured are more or less aligned with PCIe request size distributions.
In Figure~\ref{fig:bandwidth}, we show the average achieved PCIe bandwidth while executing BFS.
%
% In this experiment, we use UVM bandwidth as the reference baseline. % In the leftmost of each dataset, we also placed UVM bandwidth result as a reference.
%
We measured the maximum achievable PCIe bandwidth with \texttt{cudaMemcpy()} to be 12.3GB/s.
%We also mark the maximum PCIe bandwidth we can reach with \texttt{cudaMemcpy()} with a red dotted line in the Figure~\ref{fig:bandwidth} and it is about 12.3GB/s.\zaid{<- come back and reword}
%
Because of the page faulting overhead present in the UVM, it can only achieve PCIe bandwidth of 9GB/s.
\pname{}'s \texttt{Naive} implementation of BFS can only reach up to 4.7GB/s PCIe bandwidth. 
This is in sync with what we observed using the toy example in Figure~\ref{fig:vtune}.
With the \texttt{merge} optimization, the PCIe bandwidth utilization increased up to 11GB/s, reaching about 90\% of the peak \texttt{cudaMemcpy()} bandwidth.
With the \texttt{Merged+Aligned} optimization, we add about 0.5 to 1GB/s of additional bandwidth utilization on top of \texttt{merge} optimization in all cases.
The GU graph has the least amount of improvement from the alignment optimization among all graphs. This is because BFS on the GU graph cannot send enough number of 128-byte requests to saturate PCIe interconnect bandwidth.
By comparing Figure~\ref{fig:request_size} and Figure~\ref{fig:bandwidth}, we can clearly see the correlation between the distribution of PCIe request sizes and the achieved bandwidths in a real application, thus confirming our the analysis in $\S$~\ref{sec:zerocopy.opt}.

\begin{figure}[t]
  \centering
  \includegraphics[width=\linewidth]{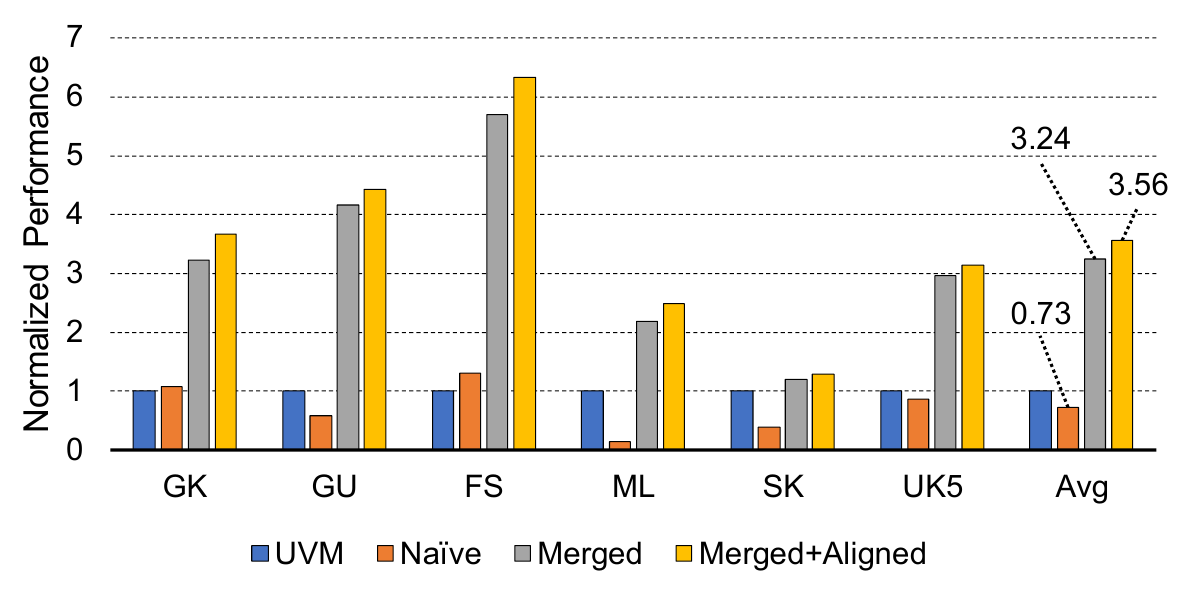}
  \vspace{-20pt}  
  \caption{BFS performance of the \texttt{Naive}, \texttt{Merged} and \texttt{Merged+Aligned} implementations against the UVM baseline.}
  \vspace{-10pt}  
  \label{fig:optperf}
\end{figure}

\begin{figure}
  \centering
  \includegraphics[width=\linewidth]{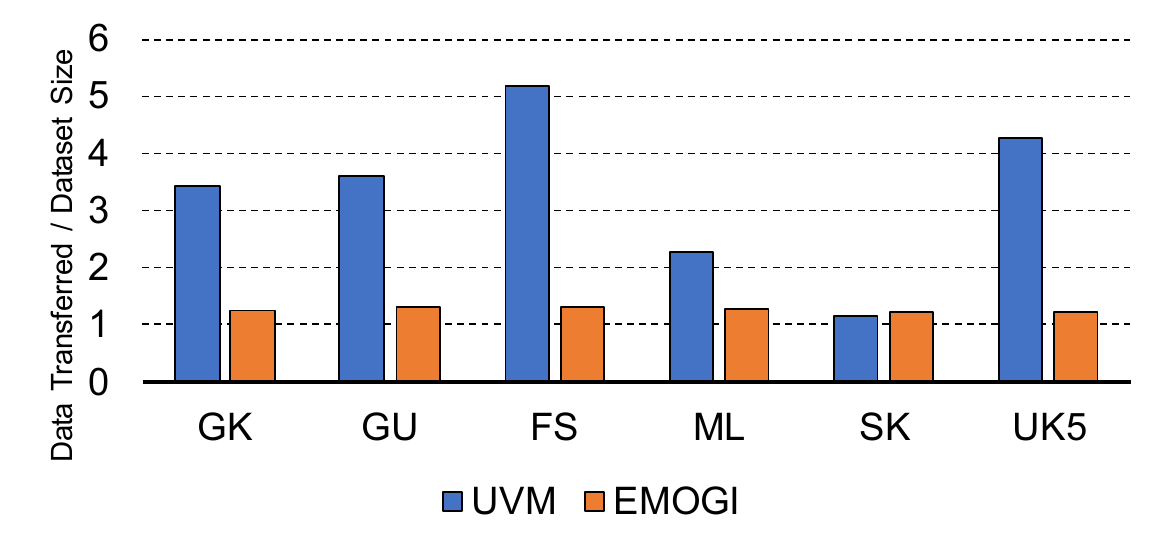}
  \vspace{-20pt}  
  \caption{I/O Read Amplification of \pname{} and the UVM baseline while performing BFS.}
  \vspace{-10pt}  
  \label{fig:oversubscription}
\end{figure}

\subsubsection{Analysis of Zero-copy Optimizations:}
\label{sec:eval.optperf}
We next evaluate the performance difference between \texttt{Naive}, \texttt{Merge} and \texttt{Merge+Aligned} implementation of BFS on various graphs and compare it with the UVM implementation.
The performance is measured based on the traversed edges per second (TEPS) and is inversely proportional to the execution time.
As shown in the Figure~\ref{fig:optperf}, the \texttt{Naive} implementation's performance is 0.73$\times$ of that of UVM on average. 
As discussed in $\S$~\ref{sec:zerocopy.opt}, this is expected as the \texttt{Naive} implementation does not use the PCIe bandwidth efficiently.
On the other hand, merging requests that go to zero-copy memory with the \texttt{Merged} implementation provides a speedup of 3.24$\times$ over the UVM baseline on average. 
For the SK graph, the performance gain using the \texttt{Merged} optimization is only 1.21$\times$  over UVM. 
This is because the SK graph can almost fit in the 16GB GPU memory. 
When we add memory access alignment optimization on top of merging of request with the \texttt{Merged+Aligned} implementation, we notice a 1.10$\times$ improvement in performance over  the \texttt{Merged} implementation on average. 
This improvement can be associated with the reduced number of PCIe requests that go out to the zero-copy memory because of the \texttt{Merged+Aligned} optimization, as was shown in Figure~\ref{fig:reqsizedetail}.

\begin{figure}[t]
  \centering
  \includegraphics[width=\linewidth]{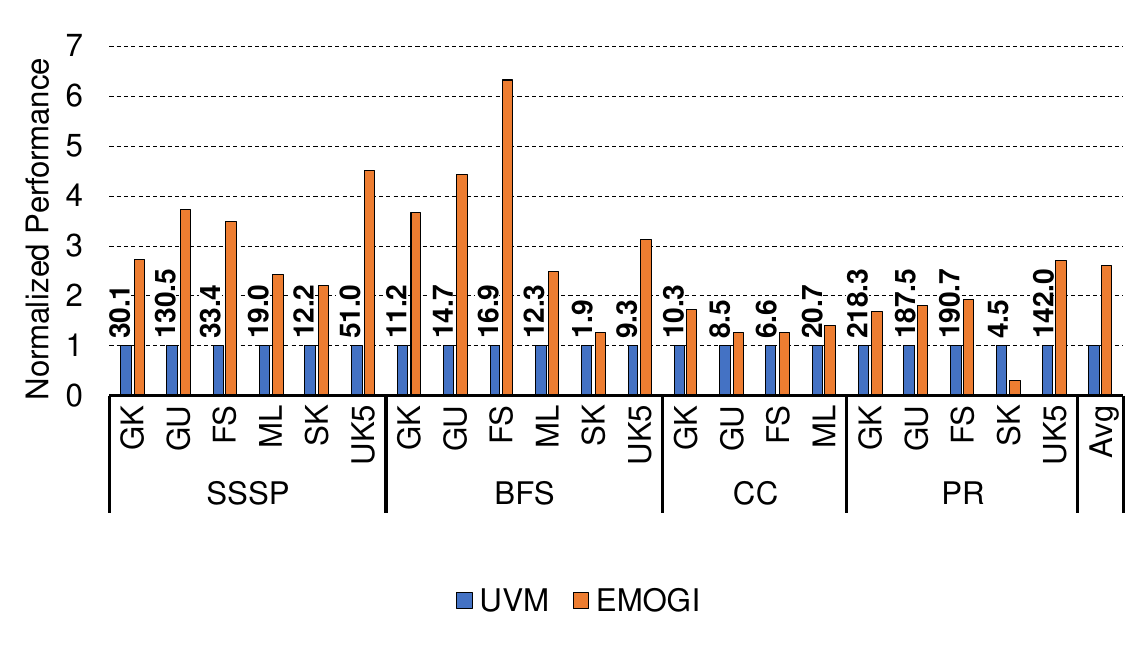}
  \vspace{-20pt}    
  \caption{Performance comparison between UVM and EMOGI with different graph traversal applications with V100. Actual execution times of \texttt{UVM} cases are written on top of the bars (in seconds).}
  \vspace{-5pt}    
  \label{fig:eval_basic}
\end{figure}

\begin{figure}[t]
  \centering
  \includegraphics[width=\linewidth]{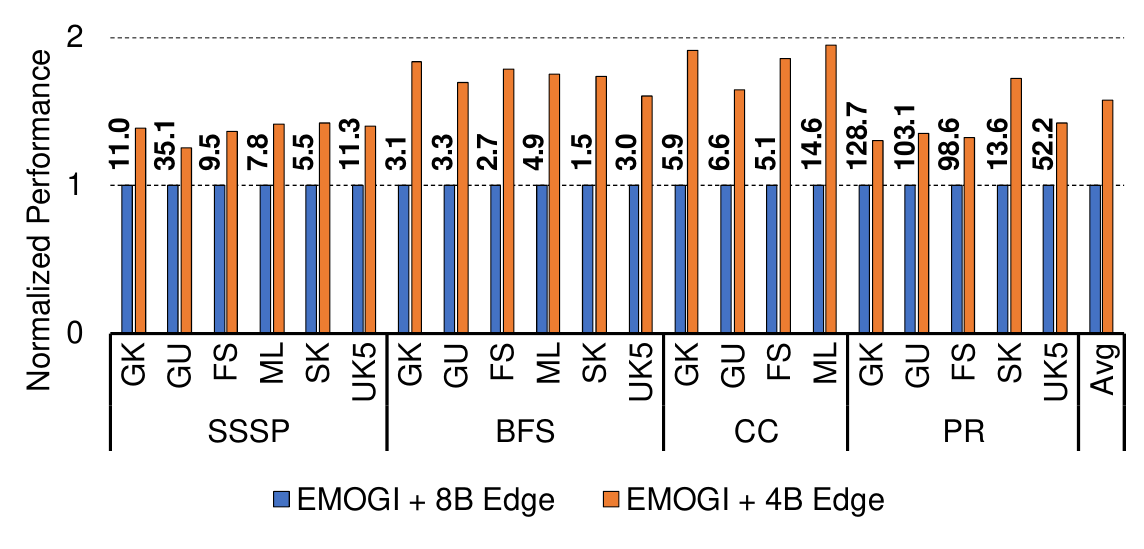}
  \vspace{-20pt}    
  \caption{Performance comparison between using 4B edge and 8B edge for EMOGI with V100. Actual execution times of \texttt{EMOGI + 8B} cases are written on top of the bars (in seconds).}
  \vspace{-5pt}    
  \label{fig:eval_4B}
\end{figure}

\subsubsection{I/O Read Amplification:}
\label{sec:ioamp}
We now demonstrate the I/O read amplification benefit of \pname{}'s fine-granular data accesses  over the 4KB page movement in UVM in BFS graph traversal. 
For this experiment, we chose the \texttt{Merge+Aligned} \pname{} implementation to represent \pname{} as it provides the best performance. 
Figure~\ref{fig:oversubscription} shows the ratio of data read from the host memory over the dataset size while performing BFS using UVM and \pname{} on each graph. 
UVM generally has a very high I/O read amplification factor, up to 5.16$\times$ for the FS graph, as for these graphs, the neighbor lists accessed during traversal are in different locations in memory and thus there is very little spatial locality exploited for each 4KB page moved.
However, the two notable exceptions to this are the ML and SK graphs as UVM's I/O read amplification factor for them is 2.28$\times$ and 1.14$\times$, respectively.
This is because the average degree of a vertex in the ML graph is 222 and the SK graph is so small that it can almost fit in GPU memory, thus making UVM's page movements a little more efficient in both cases.
In contrast, \pname{}'s I/O read amplification factor doesn't exceed 1.31$\times$.
This is because the fine-granular, merged, and aligned data access to zero-copy memory allow \pname{} to efficiently move only the necessary bytes over the slow PCIe interconnect.

\subsection{Beyond BFS}
\label{sec:moreapps}
% \subsection{Extending \pname{} to SSSP and CC}

In this section, we apply \pname{}'s optimization techniques to other graph traversal applications and measure their execution time.
%
%In this section, we expand the analysis of EMOGI to other graph traversal applications and measure the execution time of each application.
%
In addition to BFS from the previous sections, we add the single-source shortest path (SSSP), connected components (CC), and PageRank (PR) applications.
%
% The initial implementations are taken from \fixme{WHERE?}.
%
We do not evaluate the performance of CC with SK and UK5 graphs as these graphs are directed.
For PR, we do not evaluate ML graph as it is a multigraph.
%We do not run SK and UK5 in CC as strongly connected component (SCC) application is needed for the directed graphs.
%
The overall performance results are shown in Figure~\ref{fig:eval_basic}.
%
% \fixme{Removing this as it is a repeat...}
% Out of 16 results, one thing worth to mention specifically is SK from BFS.
% %
% Since the size of SK is so small that almost entire dataset fits in GPU memory, the UVM performance of SK is nearly matching EMOGI.
% %
% In case of SK, to be more realistic, should be just copied to the GPU memory using \texttt{cudaMemcpy()} and avoid any page fault overhead from UVM.
%

\pname{} provides the best performance for all the graph traversal applications and graph datasets we studied. 
On average, \pname{} is 2.60$\times$ faster than UVM.
For CC and PR, \pname{} shows relatively lower speed-ups over UVM than the other applications.
In the case of SSSP and BFS, a specific vertex is selected as a root vertex and the applications start traversing the entire graph from the root vertex.
However, with CC and PR,
instead of picking a specific vertex to start with, all vertices are set as root vertices and the entire edge list is traversed.
In this case, the application data access pattern is similar to streaming the edge list resulting in more spatial locality when compared to the other applications and less I/O read amplification for UVM.

Using smaller datatypes can reduce the overall PCIe traffic and therefore reduce the overall execution time as well.
In Figure~\ref{fig:eval_4B}, we show the performance comparison of EMOGI when using 4-byte edge list vs. 8-byte edge list.
On average, we observe about 1.57$\times$ of performance difference when using 4-byte over 8-byte for EMOGI.
In case of GK and ML graphs in CC, the performance differences are nearly 2$\times$.
The performance differences in SSSP are relatively smaller compared to the other applications since SSSP needs to transfer the weight values as well.
Due to the higher computation to memory ratio in PR~\cite{8425449}, PR also shows relatively smaller performance differences.
\subsection{Comparison with Previous Works}
In this section, we compare \pname{} with the current state-of-the-art GPU solutions for out-of-memory graph traversals, HALO~\cite{Gera20} and Subway~\cite{Sabet20}.
Due to their varying runtime requirements, we also modify our \pname{} testing environment for  accurate comparisons.
The details of the modifications are described in the following sections.
\subsubsection{HALO}
HALO proposes a new CSR reordering method to improve data locality and data transfer during graph traversal with UVM. Since the source code of the HALO is not publicly available, we compare \pname{} with the results available in the published paper.
As HALO's results were gathered using a Titan Xp GPU, we also use a Titan Xp instead of V100 for fair comparison and re-measure our execution times.
The comparison results are shown in the Table~\ref{tab:comparison_HALO}. 
Overall, \pname{} shows 1.34$\times$ to 3.19$\times$ of speedups against HALO.

\subsubsection{Subway}
Subway proposes a design of graph partitioning that preprocesses to determine the activeness of a vertex.
%Vikram: moved it here to make it HALO and Subway split.
Instead of relying on UVM, Subway focuses on generating a temporal small CSR (also called subgraph) that fits in the GPU memory in each  iteration.
Since the original CSR is located in the host memory, CPUs need to be in charge of generating the temporal CSR for every iteration.
To transfer the new temporal CSR, \texttt{cudaMemcpy()} is called by the host program.

We evaluate Subway using all the publicly available source codes (SSSP, BFS, and CC) with our platform described in $\S$~\ref{sec:eval.setup}.
%For the comparison, we use all publicly available source codes (SSSP, BFS, and CC) of Subway and evaluate them in our platform described in $\S$~\ref{sec:eval.setup}.
%
%We select Subway-async implementation for comparison with \pname{} as Subway-async provides the best performances among all available implementations.
Since one of the goals of \pname{} is to avoid any data manipulation,  
we include the subgraph generation time of Subway as well in our measurements. 
The publicly available implementation of Subway fails to execute on the GU graph due to unidentified CUDA out-of-memory errors and it cannot execute on the ML graph as the framework currently supports a maximum of $2^{32}$ edges.
%Subway uses 4-byte data type in the edge list, and therefore we re-evaluate \pname{} with the same edge list for a fair comparison.
%
The comparison results are shown in the Table~\ref{tab:comparison_subway}. 
Overall, across all the graph datasets and graph traversal algorithms, \pname{} observes speedups of 1.57$\times$ to 4.73$\times$.

\begin{figure*}
  \centering
  \includegraphics[width=\linewidth]{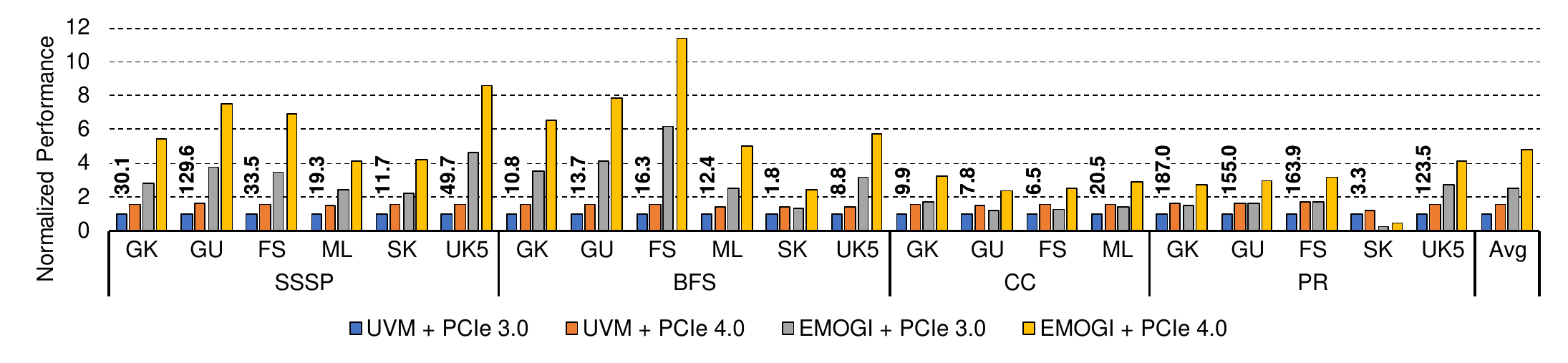}
  \vspace{-20pt}    
  \caption{Performance comparison between UVM and EMOGI using PCIe 3.0 and PCIe 4.0. All results are measured in DGX A100. Actual execution times of \texttt{UVM + PCIe 3.0} cases are written on top of the bars (in seconds).}
  \label{fig:ampere}
\end{figure*}

\subsection{Performance Scaling with PCIe 4.0}

% \fixme{Vikram lock to rewrite based on updated result.}
As was shown in $\S$~\ref{sec:pcieband} and $\S$~\ref{sec:eval.optperf}, \pname{} can nearly saturate the PCIe 3.0 bandwidth while out-performing the UVM implementation.
NVIDIA's latest GPU, the Ampere A100, communicates with the host memory over the PCIe 4.0 interconnect.
PCIe 4.0's measured peak bandwidth, %at
approximately 24GB/s, is twice as much as PCIe 3.0's peak measured bandwidth of approximately 12GB/s.
In this section, we study the ability of both UVM and \pname{} to take advantage of the increased bandwidth in accessing the host memory.
To this end, we use a DGX A100 machine ~\cite{dgxa100dataset} with the A100 GPU and Dual AMD Rome 7742 CPUs paired with 1TB of system memory.
This machine allows us to switch the root port to run in either PCIe 3.0 mode or PCIe 4.0 mode.
Neither the \pname{} implementation nor the UVM implementation was re-optimized for the A100 GPU in these experiments.
A100 memory is throttled to 16GB.

The overall evaluation results comparing the performance of UVM and \pname{} on the DGX A100 system are shown in Figure~\ref{fig:ampere}.
Here, we normalize the performance speed-up achieved by each configuration to the UVM implementation running on the A100 GPU with the PCIe 3.0 interconnect.  
While \pname{}'s performance scales by 1.88$\times$ on average with the faster interconnect, UVM's performance scales by only 1.53$\times$ on average.
This is because the UVM implementation suffers from page fault handling overhead when accessing pages of the edge list in host memory. 
The page fault handler is part of the UVM driver running on the CPU and can't keep up to make use of the higher bandwidth of the PCIe 4.0 interface.
However, \pname{} doesn't suffer any page faulting overhead as the edge list is pinned in host memory, leading to \pname{}'s performance scaling almost linearly with the PCIe bandwidth.

\section{Discussion}
\label{sec:discuss}

\textbf{Extending to other input formats:} In this paper, EMOGI is targeting CSR which is used by many popular graph processing frameworks~\cite{gunrock,galois,nvGRAPH,graphbig,ligra}, but the main idea of EMOGI can be extended to different formats as well.
The most immediate applicable format is compressed sparse column (CSC).
The edgelists in CSR represent outgoing edges (push-based), but the edgelists in CSC represent incoming edges (pull-based).
Although the directions are different between the CSR and CSC, the memory access pattern to the edgelists in both input formats is identical.
Another interesting format expansion for EMOGI would be dynamic graphs~\cite{sha17} and compressed graphs~\cite{sha19, kaczmarski2015improving}.
The graph input formats used in these works are not strictly identical to the classical CSR format, but their fundamental structures resemble CSR to retain some level of data locality for better bandwidth utilization.
Therefore, the EMOGI's zero-copy memory access optimization strategies can be applied to these formats as well.

\begin{table}[t]
  \caption{Execution time comparison with HALO~\cite{Gera20}. NVIDIA Titan Xp (12GB) used.}
  \label{tab:comparison_HALO}
  \begin{tabular}{ccccc}
    \toprule
    \multirow{2}{*}{Application} & \multirow{2}{*}{Graph} & \multicolumn{2}{c}{Exe. Time} & \multirow{2}{*}{Speedup}\\
    \cline{3-4}
                &       & HALO & EMOGI &\\
    \midrule
    \multirow{4}{*}{BFS}    & ML  & 9.54s  & 4.43s & 2.15$\times$\\
                            & FS  & 8.27s  & 2.59s & 3.19$\times$\\
                            & SK  & 2.17s  & 1.62s & 1.34$\times$\\
                            & UK5 & 6.03s  & 4.00s & 1.51$\times$\\
    \bottomrule
  \end{tabular}
\end{table}

\begin{table}[t]
  \caption{Execution time comparison with Subway~\cite{Sabet20}. NVIDIA Tesla V100 (16GB) used. 4-byte edge used due to the Subway requirement.}
  \label{tab:comparison_subway}
  \begin{tabular}{ccccc}
    \toprule
    \multirow{2}{*}{Application} & \multirow{2}{*}{Graph} & \multicolumn{2}{c}{Exe. Time} & \multirow{2}{*}{Speedup}\\
    \cline{3-4}
                &       & Subway & EMOGI &\\
    \midrule
    \multirow{4}{*}{SSSP}   & GK  & 20.96s & 7.94s & 2.64$\times$\\
                            & FS  & 14.95s & 6.97s & 2.14$\times$\\
                            & SK  & 8.99s  & 3.92s & 2.30$\times$\\
                            & UK5 & 25.78s & 8.08s & 3.19$\times$\\
    \midrule
    \multirow{4}{*}{BFS}    & GK  & 6.88s & 1.66s & 4.14$\times$\\
                            & FS  & 4.22s & 1.49s & 2.83$\times$\\
                            & SK  & 1.69s & 0.85s & 1.99$\times$\\
                            & UK5 & 8.75s & 1.85s & 4.73$\times$\\
    \midrule
    \multirow{2}{*}{CC}     & GK  & 6.34s & 3.11s & 2.04$\times$\\
                            & FS  & 4.31s & 2.75s & 1.57$\times$\\
    \bottomrule
  \end{tabular}
\end{table}

\textbf{Additional optimizations:} There are several additional optimizations available for EMOGI such as data compression and data caching in GPU global memory.
Data compression on graph~\cite{sha19, kaczmarski2015improving} can reduce the total amount data transferred to GPU and we can obtain a similar effect to increasing the external interconnect bandwidth.
As discussed in $\S$~\ref{sec.emogi.opt}, GPU is severely underutilized due to the low external interconnect bandwidth. Thus the idling GPU cores can be potentially used to decompress data from the host memory, without interfering with the original graph traversing process.
For data caching, a work similar to \cite{zhang17, lakhotia17} can be applied to exploit data locality further.
For this optimization, we expect that a workload with high vertex revisits, such as PR, would benefit the most.
However, one thing to note is that currently there is no full hardware-based mechanism to naturally use the GPU global memory as a large cache and therefore a software-based caching mechanism needs to be implemented.
\section{Related Works}
\label{sec:relatedwork}
\textbf{Graph Analytics on CPU:}
Efficient graph traversal algorithm implementation on CPUs have been extensively studied in the past\cite{pearce10,GoFFish,graphmat,Graphicionado,spark, galois, ligra}. 
% There are multiple graph traversal works implemented for CPU.
%
We note that from these studies two popular frameworks Galois~\cite{galois} and Ligra~\cite{ligra} have emerged. 
%Two popular frameworks are Galois~\cite{galois} and Ligra~\cite{ligra}.
%
To compensate the low parallelism of CPU, these frameworks support the workload scaling to multi-CPU systems such as non-uniform memory access (NUMA) nodes. 
% Scale-out and MPI parallelism is enabled in these frameworks to compensate the low parallelism of the CPUs. 
%
However, the focus of these prior work has been on efficient graph partition and data allocation schemes across different nodes to minimize the node-to-node data movements.
% For such works, the focus is on optimizing a graph partition and data allocation among different nodes to minimize the node to node data transfer.
%
Therefore, it is inevitable to do a pre-processing for efficient data organization. 
Beside the frameworks, there are other numerous works \cite{pearce10,GoFFish,graphmat,Graphicionado,spark} based on CPU.

\textbf{Graph Analytics on GPU:}
Graph traversal algorithms such as BFS exhibit a massive amount of parallelism. 
%Graph traversal algorithms such as BFS, SSSP and CC exhibit massive amount of parallelism. 
This has led to increasing research in leveraging the massive computation power offered by GPUs to speed up graph analytics. 
Prior work focused on improving the performance of graph traversal algorithms either by making GPU specific algorithmic improvements~\cite{xbfs,maxwarp,cusha,medusa, gunrock,digraph, scalablegraphtraversal} or by performing data transformations~\cite{Tigr, compiler}. 
However, most of these works assume graphs fit in the GPU memory. 

Practical graphs, on the other hand, often cannot fit into the GPU memory.
Web graphs~\cite{BoVWFI,BRSLLP}, social network graphs~\cite{Friendster} and bio-medical graphs~\cite{GAP} can be significantly larger than available GPU memory (see Table~\ref{tab:dataset}). 
To address this, prior works have proposed either to partition the input graph and loading only those edges that are needed during computation~\cite{gunrock,graphreduce,graphie,Sabet20} or leveraging automatic memory oversubscription using UVM~\cite{chai,Gera20,kokkos, etc, batchaware, graphchallenge18, graphchallengetc19}. 
%Among several input graph partitioning schemes ~\cite{gunrock,graphreduce,graphie,Sabet20}, the notable ones are GraphReduce and Subway~\cite{Sabet20}. 
For example, GraphReduce~\cite{graphreduce} partitions the oversized graphs and does explicit memory management between the GPU and the host memory.  
Recently, Subway~\cite{Sabet20} further improved the design of the input partitioning scheme using GPU-accelerated subgraph generation pre-processing technique that tracks activeness of a vertex and also by generating subgraphs asynchronously. 
%\pname{} differs from these works as 
\pname{} does not perform any explicit memory management or pre-processing of the graph. 

Alternatively, to support large graphs in GPU, programmers can use UVM which does automatic memory oversubscription~\cite{pascalwhite, voltawhite}. 
%Once the graph is loaded to the host memory, the driver and the runtime hardware does on-demand page migrations transparently between the host and the GPU.
Prior works such as~\cite{chai,Gera20,kokkos, etc, batchaware, graphchallenge18, graphchallengetc19, pageplacement, mosaic, gpuswap} have observed significant overhead from UVM and have proposed optimizations such as overlapping IO and computation~\cite{chai}, memory spaces\cite{kokkos}, memory throttling~\cite{etc}, modifying driver to support larger page fault batch sizes~\cite{batchaware} and reordering of graphs to enhance locality in UVM~\cite{Gera20}. 
%\pname{} is along the same direction as the above proposed systems. 
Instead of leveraging the previously proposed optimizations, \pname{} takes a step back and revisits the reasoning behind the performance degradation with UVM.
Like~\cite{pciebwlow}, \pname{} initially observes the PCIe bandwidth utilization being low for graph traversal applications.
As shown in $\S$~\ref{sec:eval}, by carefully re-orchestrating the memory access pattern using direct access, \pname{} is able to boost graph traversal execution performance for large graphs without any additional optimizations. 
Indeed the prior proposed software and hardware optimizations can be exploited by \pname{}. 
% and in-turn be merged into UVM to optimize it for graph applications. 
We leave this as future research.  Also, \pname{} could be easily incorporated into a library to lessen the programmer's effort and provide out-of-the box performance improvements. 
%future UVM implementations as a way to bypass the paging mechanism for data with little short-term temporal reuse.
%\fixme{Add selling point of E-UVM here.} 

\textbf{Multi-GPU and Collaborative CPU-GPU:} %this is added to cite VLDB reviewers :) 
Aside from single GPU graph traversal, prior works have proposed using multi-GPU ~\cite{multigpugraph,medusa,scalblesimd,gunrock} and collaborative CPU-GPU computation to meet the needs of large graphs computation~\cite{graphchallenge18,chai,garaph,yokechicken,thinkvertexthinkgraph}.
Multi-GPU and collaborative CPU-GPU computing are enabled using UVM where hardware moves the pages on-demand across different computing blocks. 
%Although in this work we do not explore these use-cases,
\pname{} can be extended to support both multi-GPU and hybrid CPU-GPU computing and we leave it as future  research.

\textbf{Architectural support for improving UVM:}
%\fixme{vikram: do we need this sub-section?} section to make some common arch people happy
Besides algorithm and system-level changes, prior works also propose hardware changes that can enable executing graph traversal algorithms efficiently on large graphs. 
Specifically, memory compression techniques to reduce the memory footprint in the GPU~\cite{etc}, efficient migration policies using hardware counters, and optimized prefetching schemes~\cite{pageplacement, mosaic, hardwareprefetcher}, and software-hardware co-design using memory hints are proposed~\cite{voltawhite,amperewhite}. 
These techniques are orthogonal to \pname{} and can be leveraged by \pname{} to gain further performance improvements in future GPU architectures.

\section{Conclusion}
\label{sec:conclusion}
In this work, we present \pname{}, a new method for optimizing the traversal of very large graphs with a GPU using zero-copy.
%presented the motivation, design, and evaluation of
Using thorough analysis of fine-grained GPU memory access patterns over PCIe to zero-copy memory, we identified key optimizations to best utilize bandwidth to zero-copy memory: merged and aligned memory accesses.
We applied these optimizations to key graph traversal applications to enable efficient GPU traversal of graphs that do not fit in GPU memory.
%including breadth-first search, single-source shortest path, and connected components
%EIMAN-6-1-20: I think the immediate sentence that follows can be made better. 
%It would be more insightful to say:
%Our experiments showed that \pname{} out-performs the state-of-the-art solutions for traversing large graphs by up to 4.73$\times$. This is because EMOGI...
%fill the ... with something insightful about how ``merged and aligned fine-grained accessses'' result in the performance improvement. Just saying EMOGI ``efficiently'' uses them and repeating the name of the optimizations and zero copy memory which you've already said a couple of times above sounds fluffy. Add some insight where I put the ... above and break the sentence down so its not as long.
%Experiments showed that \pname{} out-performs the state-of-the-art solutions for traversing large graphs by up to 4.73$\times$ as it efficiently uses the available bandwidth with merged and aligned fine-grained accesses to zero-copy memory. Compared to UVM-based traversal code,  \pname{} substantially reduces I/O read amplification and boosts data access efficiency.
Our experiments show that \pname{} out-performs the state-of-the-art solutions for traversing larges graphs.
This is because \pname{} avoids I/O read amplification by leveraging efficient fine-grained accesses to fetch only the needed bytes from zero-copy memory.
Furthermore, \pname{}'s performance scales almost linearly with the improved bandwidth of newer interconnects as it is not bottle-necked by the page fault handling overhead of traditional methods using UVM.

%\vspace{20ex}

%\begin{acks}
% This work was supported by the [...] Research Fund of [...] (Number [...]). Additional funding was provided by %[...] and [...]. We also thank [...] for contributing [...].
%\end{acks}

\begin{acks}
This work was partially supported by the Applications Driving Architectures (ADA) Research Center and Center for Research on Intelligent Storage and Processing-in-memory (CRISP), JUMP Centers co-sponsored by SRC and DARPA, IBM-ILLINOIS Center for Cognitive Computing Systems Research (C3SR). This work would not have been possible without the generous hardware donations from Xilinx and NVIDIA. We also thank the IMPACT members, anonymous reviewers, and a shepherd for their constructive feedbacks, and the chief editors' coordination in the shepherding process.
\end{acks}

%\clearpage
\balance
\bibliographystyle{ACM-Reference-Format}
\bibliography{ref}

\end{document}